\documentclass[10pt,letterpaper]{article}
\usepackage{amsmath}
\usepackage{amsfonts}
\usepackage{amssymb}
\usepackage{url}
\usepackage{graphicx}
\usepackage{dcolumn}
\usepackage{bm}
\usepackage{hyperref}
\usepackage{authblk}
\usepackage{upgreek}
\usepackage{booktabs}
\usepackage{natbib}

\makeindex      

\title{EXONEST: The Bayesian Exoplanetary Explorer}

\author[1]{Kevin H. Knuth}
\author[2]{Ben Placek}
\author[3]{Daniel Angerhausen}
\author[1]{Jennifer L. Carter}
\author[1]{Bryan D'Angelo}
\author[1]{Anthony D. Gai}
\author[1]{Bertrand Carado}

\affil[1]{Dept. of Physics, University at Albany, Albany, NY 12222, USA}
\affil[2]{Dept. of Sciences, Wentworth Institute of Technology, Boston, MA 02115, USA}
\affil[3]{Center for Space and Habitability, University of Bern, Bern 3012, Switzerland\\ Blue Marble Space Institute of Science, Seattle, WA 98154, USA}

\date{20 October 2017}
\begin{document}
\maketitle

\abstract{The fields of astronomy and astrophysics are currently engaged in an unprecedented era of discovery as recent missions have revealed thousands of exoplanets orbiting other stars. While the Kepler Space Telescope mission has enabled most of these exoplanets to be detected by identifying transiting events, exoplanets often exhibit additional photometric effects that can be used to improve the characterization of exoplanets. The EXONEST Exoplanetary Explorer is a Bayesian exoplanet inference engine based on nested sampling and originally designed to analyze archived Kepler Space Telescope and CoRoT (Convection Rotation et Transits plan\'etaires) exoplanet mission data. We~discuss the EXONEST software package and describe how it accommodates plug-and-play models of exoplanet-associated photometric effects for the purpose of exoplanet detection, characterization and scientific hypothesis testing. The current suite of models allows for both circular and eccentric orbits in conjunction with photometric effects, such as the primary transit and secondary eclipse, reflected light, thermal emissions, ellipsoidal variations, Doppler beaming and superrotation. We~discuss our new efforts to expand the capabilities of the software to include more subtle photometric effects involving reflected and refracted light.
We discuss the EXONEST inference engine design and introduce our plans to port the current MATLAB-based EXONEST software package over to the next generation Exoplanetary Explorer, which will be a Python-based open source project with the capability to employ third-party plug-and-play models of exoplanet-related photometric effects.}

\pagebreak
\section{Introduction}

We are currently enjoying an unprecedented era of exploration and discovery. July of 2015 saw the New Horizons probe's fly-by of Pluto and Charon, which marked the end of mankind's initial exploration of the solar system. We are now beginning to explore the neighboring star systems by discovering and characterizing their planets (exoplanets). As stated in the 2014 NASA Strategic Plan, ``We are navigating a voyage of unprecedented scope and ambition: seeking to discover and study planets orbiting around other stars and to explore whether they could harbor life'' \cite{NASA:2014}. To date, we have identified 4496 exoplanet candidates of which 3502 are confirmed planets in over 1648 star systems \cite{ExoplanetArchive:2017}.

One of the most successful missions to date is the Kepler Space Telescope (Kepler), which~was designed to monitor the light intensity (photometry) from approximately 150,000 stars in the constellations of Cygnus and Lyra \cite{Borucki:2010:Kepler}. While the characterization of exoplanets in terms of the chemistry and the detection of life will require a serious commitment to spectroscopic studies, photometry will continue to play an important role in exoplanet detection and characterization. Future surveys like the Transiting Exoplanet Survey Satellite (TESS) \cite{Ricker+etal:2010:TESS, Placek+Knuth+Angerhausen:2016:Kepler+TESS} will pave the way for higher precision photometry missions like the James Webb Space Telescope (JWST) \cite{Beichman+etal:2014:JWST, Placek+Angerhausen+Knuth:2017:optimization}, the CHaracterising ExOPlanets Satellite (CHEOPS) \cite{CHEOPS, CHEOPSMHP, Broeg+etal:2013, Placek+Angerhausen+Knuth:2017:optimization}, the PLAnetary Transits and Oscillations (PLATO) mission \cite{PLATO, PlatoSciReq, Hippke+Angerhausen:2015} and the Wide Field Infrared Survey Telescope (WFIRST) \cite{WFIRST}. Increasingly accurate representations of photometric effects will become more important with these higher precision missions.

In this paper, we summarize the EXONEST software package for detecting and characterizing exoplanets \cite{Placek+Knuth+Angerhausen:EXONEST, Placek:thesis:2014}, as well as introduce our new efforts toward more careful modeling of reflected light, refracted light and atmospheric effects and their incorporation into the next generation of the Exoplanetary Explorer software package. We are currently focused on porting and expanding the MATLAB-based EXONEST software package into the next generation Exoplanetary Explorer software package, which will accommodate multiple planet systems, three-body and multi-body orbital mechanics, as well as more subtle photometric effects. The Exoplanetary Explorer will be a Python-based open source project with the capability to employ third-party plug-and-play models of exoplanet-related photometric effects.

\section{EXONEST}

\begin{figure}[t]
  \centering
  \includegraphics[width=0.7\textwidth]{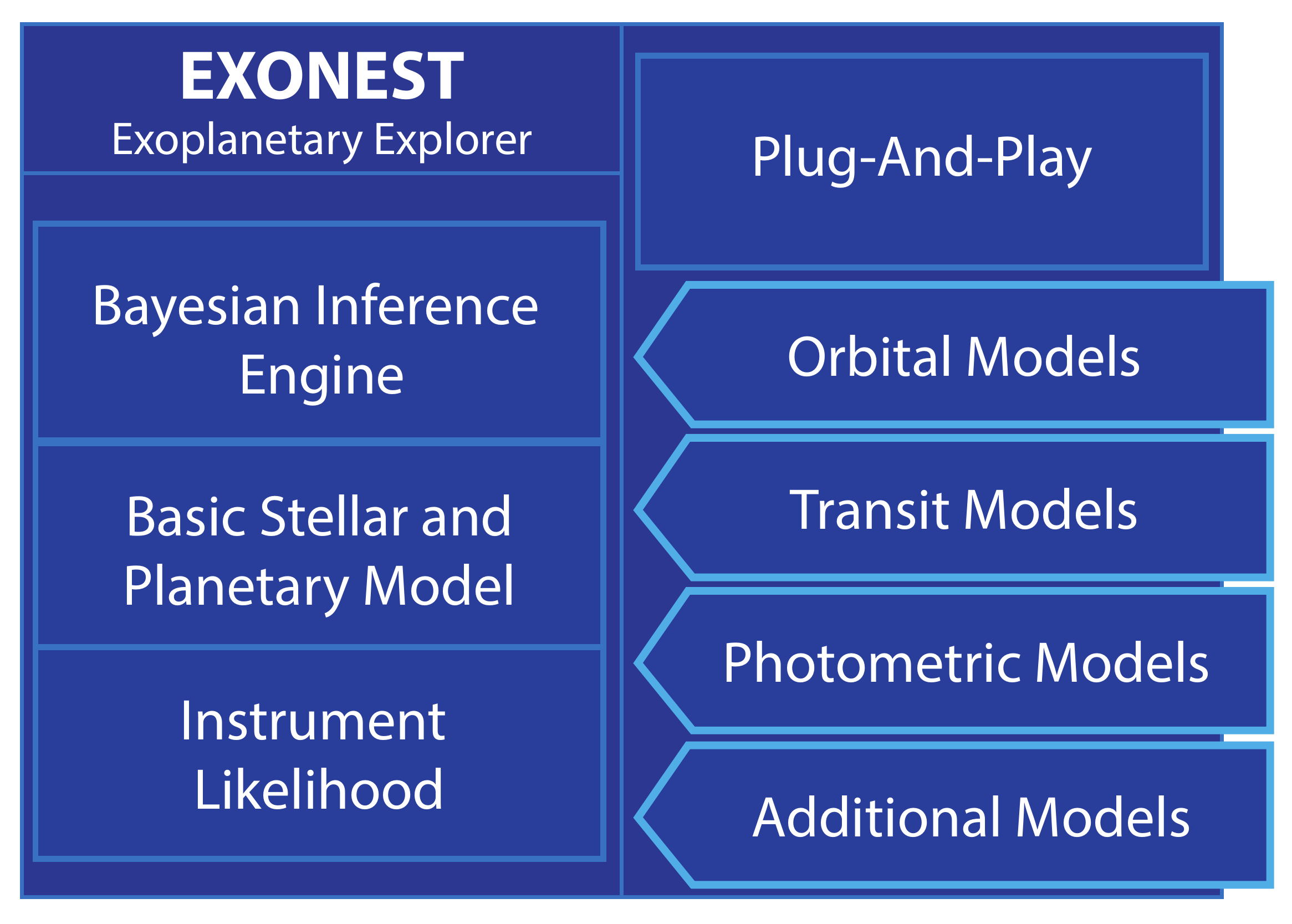}
  \caption{An illustration of the structure of the EXONEST software package.} \label{fig:exonest-diagram}
\end{figure}

The EXONEST software package (Figure \ref{fig:exonest-diagram}) was originally developed as a part of Ben Placek's Ph.D. thesis \cite{Placek+Knuth+Angerhausen:EXONEST, Placek:thesis:2014}. Written in MATLAB, EXONEST consists of a core set of stellar and planetary models, a nested sampling-based Bayesian inference engine that can utilize the original nested sampling algorithm \cite{Skilling:2004:nested, Skilling:2006:nested, Sivia&Skilling}, the MultiNest variant \cite{Feroz+etal:2009, Feroz+etal:2011a, Feroz+etal:2013} and the Metropolis--Hastings Markov chain Monte Carlo (MCMC) sampling algorithm \cite{Metropolis:1953, Hastings:1970}. EXONEST can also accept a set of plug-and-play models, which consist of orbital models, instrument likelihood functions, transit models, a set of photometric models and any additional user-defined models. This is currently implemented by modifying the code by calling one or more subroutines. In the future, these options will be selected, or unselected, via a run file and/or a graphical user interface.

EXONEST incorporates an efficient orbit integrator, as well as detailed models for four photometric effects: reflected light, thermal emissions, Doppler boosting or beaming and ellipsoidal variations of the host star \cite{Placek+Knuth+Angerhausen:EXONEST, Placek:thesis:2014}. Both likelihood functions and priors on the parameter values can be~specified.

In the following sections, we will summarize the orbital and photometric models employed by EXONEST and our current work to extend and improve them. We will then conclude by discussing the Bayesian inference engine and the need for improvement.

\section{Orbital Models}

EXONEST incorporates an efficient orbit integrator that can accommodate both circular and elliptical orbits \cite{Brown:photometric:2009}, known as Keplerian orbits, producing both the star-planet distance $r(t)$ and the orbital phase $\theta(t)$ as a function of time. The current implementation of EXONEST is designed to accommodate one exoplanet per system so that only one period of the orbit need be integrated. The~data are phase-wrapped to decrease computation time.

\begin{figure}[t]
  \centering
  \includegraphics[width=0.93\textwidth]{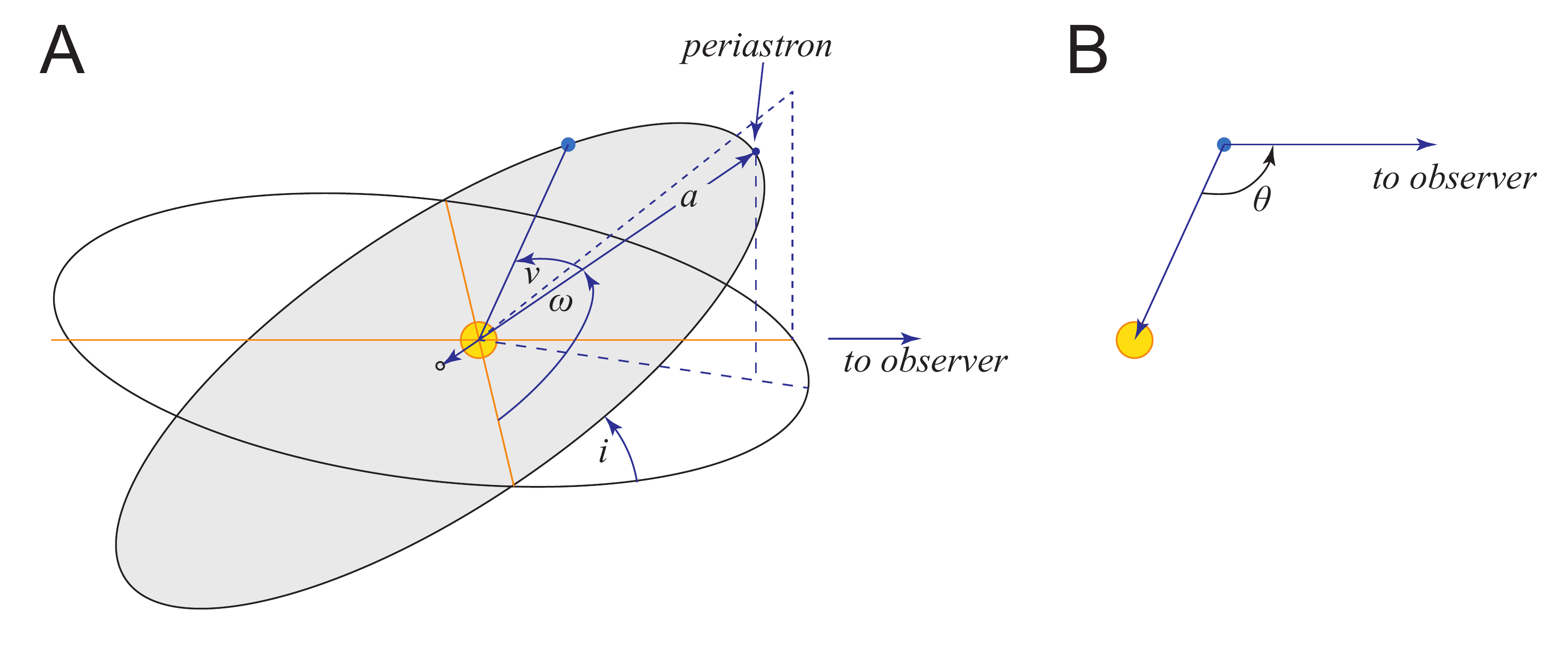}
  \caption{(\textbf{A}) An illustration of the anatomy of a typical Keplerian orbit. The planet moves on an elliptical orbit with the star located at one focus. The semi-major axis, $a$, of the ellipse is used to characterize the star-planet distance. The argument of the periastron, $\omega$, is the angle describing how the periastron (point of closest approach) is rotated with respect to the observer. The true anomaly, $\nu$,~describes the position of the planet along the orbit in the frame of the host star with the origin at the periastron. (\textbf{B}) The angle $\theta(t)$ indicates the orbital phase, which is defined as the angle between the planet-star vector and the planet-Earth vector. } \label{fig:orbital-elements}
\end{figure}

The basic anatomy of a Keplerian orbit is illustrated in Figure \ref{fig:orbital-elements}. The star is situated at the focus of the elliptical orbit. The {{periastron}} (point of closest approach) may be rotated by an angle $\omega$, which~is called the {{argument of the periastron}}, and the position of the planet along the orbit is designated by the angle from periastron, $\nu$, which is called the {{true anomaly}}. The~length of the semimajor axis, $a$,~is used to characterize the star-planet distance, $r(t)$, which varies as a function of the true anomaly according~to:
\begin{equation} \label{eq:eccentric-orbits}
r(t) = \frac{a (1-e^2)}{1 + e\cos{(\nu(t))}} \; ,
\end{equation}
where $e$ is the eccentricity of the orbit.

The upcoming re-write of the EXONEST code will accommodate multiple exoplanets in different orbits. This is accomplished by integrating a single period of each exoplanet orbit and labeling the time series of the data in terms of the orbital phase of each orbit.

A full three-body orbital model has been implemented for EXONEST. The model can handle cases where the planets do or do not orbit in a common plane. Despite its utility, the model is computationally expensive as the initial conditions of each planet must be fully parameterized, thus~significantly increasing the number of parameters to be estimated. In addition, orbits in three-body situations are, in general, not periodic. Unlike two-body periodic elliptical (Keplerian) orbits, which need only be integrated over a single period to make predictions for the entire dataset, three-body orbits must be integrated over the entire length of the dataset, dramatically increasing the computations necessary to generate the predicted photometric flux.

\section{Photometric Effects}

In this section, we consider the main photometric effects that are detectable with instruments exhibiting precision ranging from 10 parts per million to 30 parts per million. These photometric effects include both planetary components and stellar components. We will begin with the planetary components by discussing transits, which produce relatively large photometric signals. This is followed by a discussion of reflected light, thermal emissions, refracted light, and atmospheric effects. We then conclude with the stellar components, which consist of Doppler boosting and ellipsoidal variations.

\begin{figure}[t]
  \centering
  \includegraphics[width=0.85\textwidth]{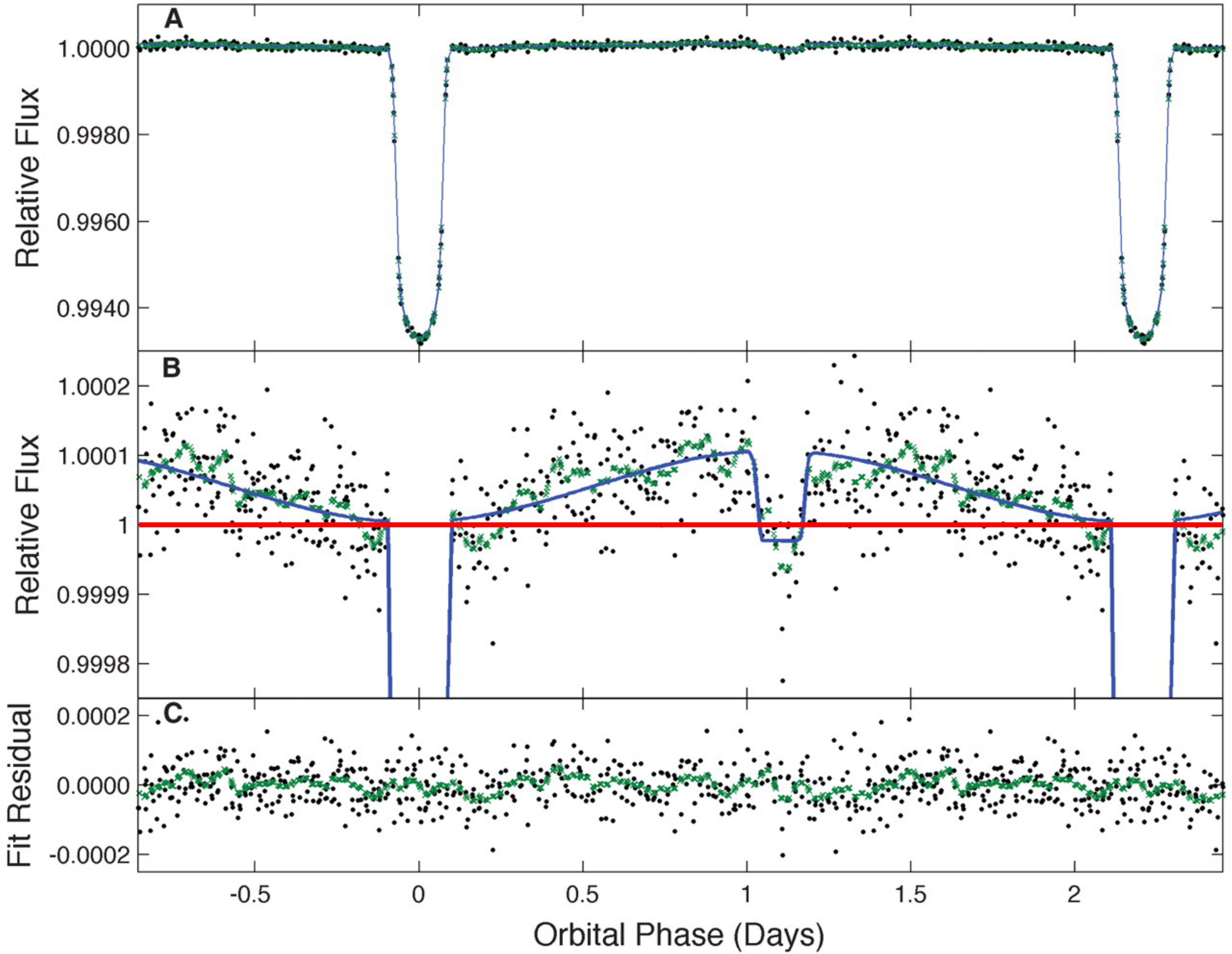}
  \caption{(\textbf{A}) Light curve of HAT-P-7b recorded by Kepler as processed and published by \mbox{Borucki et al.~\cite{Borucki-etal:2009:Science}} (reprinted here with permission). (\textbf{B}) The same light curve at greater magnification.  The red line marks the baseline flux. The plot in (\textbf{C}) illustrates the residual, which is the difference between the data and the model. The lack of obvious structure in the residual indicates that most of the effects have been well modeled. }
 \label{fig:HAT-P_7b}
\end{figure}

Figure \ref{fig:HAT-P_7b} illustrates the recorded data and light curve fit from the exoplanet HAT-P-7b \cite{Borucki-etal:2009:Science}. The~primary and secondary eclipses are clearly seen at orbital phase zero days and 1.1 days,
respectively. In Figure \ref{fig:HAT-P_7b}B, planetary flux variations, such as reflected light, are evident in the sinusoidally-varying flux and in the presence of the secondary eclipse as the planet passes behind the host star. The fact that the flux during the secondary eclipse dips below the baseline (red line) indicates that the planet is also emitting a great deal of thermal radiation, which is hidden as the planet passes behind the host star.

\subsection{Transits and Eclipses}

\begin{figure}[t]
  \centering
  \includegraphics[width=0.45\textwidth]{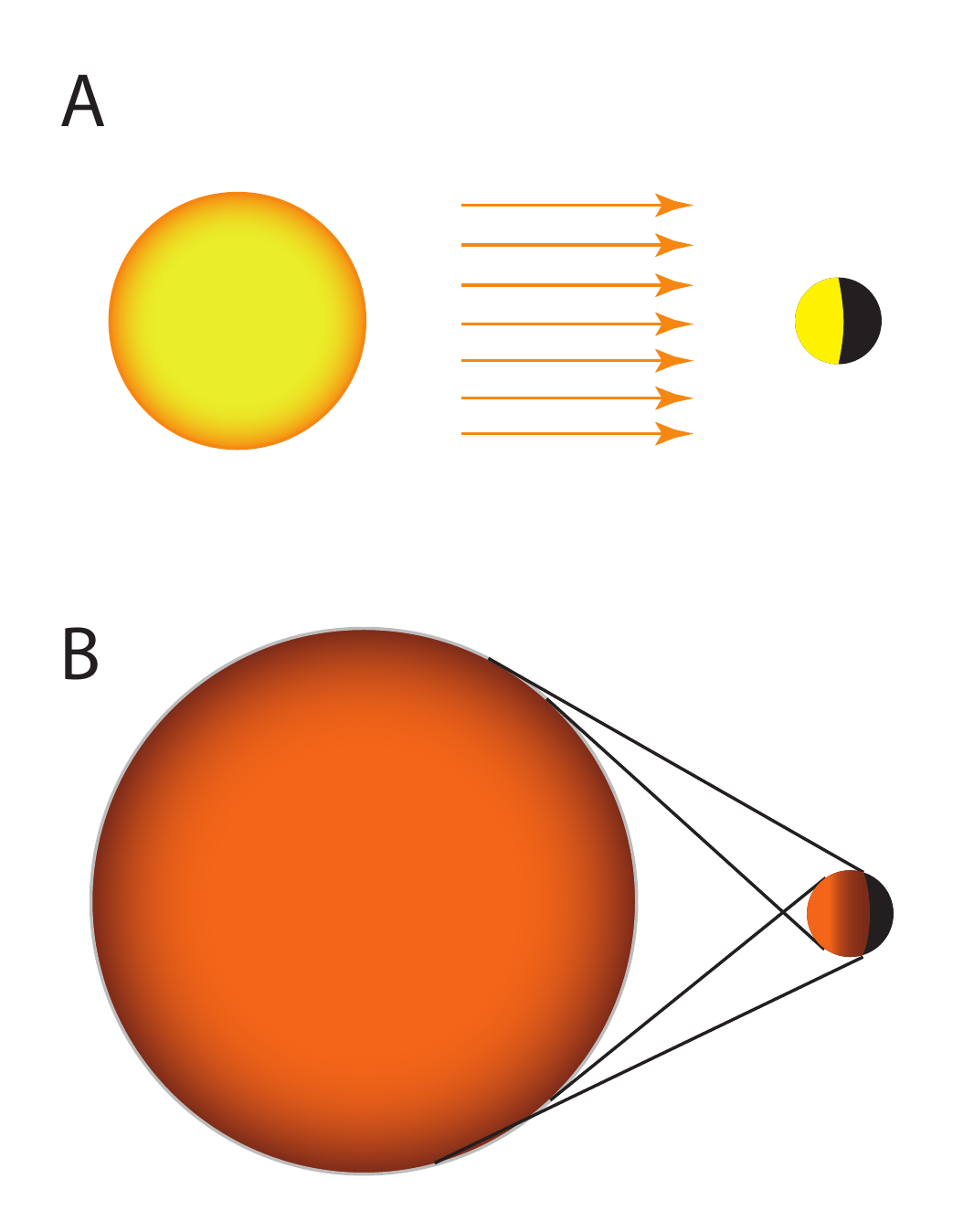}
  \caption{The range of geometries of a planet illuminated by a star. (\textbf{A}) A planet orbits an average star at a great distance. The star's rays are more or less parallel, so that one-half of the planet is in illuminated daylight and the other half is in night. (\textbf{B}) A planet in a close-in orbit around a giant star is more than 50\% illuminated with a full daylight zone, a penumbral zone and a night zone.}
  \label{fig:illuminated-planets}
\end{figure}

The most prominent photometric effect is known as the transit or {{primary transit}}, which occurs when the planet passes in front of the host star blocking a fraction of the starlight \cite{Seager+Mallen:2003}.
The primary transit of the exoplanet HAT-P-7b can be seen in Figure \ref{fig:HAT-P_7b} at an orbital phase of zero days and again at 2.2~days, which is its orbital period. The bottoms of the primary transits are typically rounded
because
the light emitted by the star is not uniform across the stellar disk, but is darker near the limbs as one is looking through the cooler upper regions of the photosphere \cite{Mandel+Agol:2002}. This effect, known as {{limb darkening}}, is illustrated by the coloration of the stars on the left side of Figure \ref{fig:illuminated-planets}A,B.
The~transit depth is proportional to the ratio of cross-sectional areas of the planet and star:
$$
\frac{{R_p}^2}{{R_{S}}^2} \; ,
$$
where $R_p$ is the radius of the planet and $R_{S}$ is the radius of the star.
Transits are relatively rare since the planet must pass between the stellar disk and the observer. Borucki and Summers found the probability that a distant observer could witness an exoplanetary transit to be $\frac{R_{S}}{a}$ \cite{Borucki+Summers:1984}.~This expression was found by considering the solid angle subtended by the exoplanet's shadow in a circular orbit. Here, we consider eccentric orbits with eccentricity $e$, and~more-or-less follow the derivation by Barnes \cite{Barnes:2007}, taking care to rigorously treat the probability distribution associated with the  true anomaly, the argument of the periastron and the inclination.

To find the probability of a transit, we first consider circular orbits and work out the prior probability of observing a planet at a given angular position defined by the true anomaly and the inclination, $(\nu, i)$. It is expected that there is no preferred orientation for the inclination of a planet's orbital plane. Moreover, for a circular orbit, the planet orbits at a constant speed so that the prior probability of the true anomaly is also constant. As a result, the joint prior probability is constant, $Pr(\nu, i | I) = \frac{1}{C}$, where it should be noted that in our expressions for the probability, $Pr$, we adopt the convention of Jaynes \cite{Jaynes:Book} and Sivia and Skilling \cite{Sivia&Skilling} in which all probabilities are written as being conditional on all of the prior information $I$ about the problem:

\begin{equation}
\begin{array}{ll}
1  = \int_0^{\uppi} di \, \sin{(i)} \; \left(\int_0^{2\uppi} { Pr( \nu, i | I ) \, d\nu }\right) \vspace{4pt}
 \\
 \hspace{8pt}= \frac{1}{C} \int_0^{\uppi} di \, \sin{(i)} \; \left(\int_0^{2\uppi} { d\nu }\right) \vspace{4pt} \\
 \hspace{8pt}= \frac{2\uppi}{C} \int_0^{\uppi} di \, \sin{(i)}\vspace{4pt}  \\
\hspace{8pt} = \frac{4\uppi}{C}
\end{array}
\end{equation}
so that:
\begin{equation} \label{eq:joint-angle-prior}
Pr( \nu, i | I) = \frac{1}{4\uppi}.
\end{equation}

In the case of eccentric orbits, one must also consider the argument of the periastron $\omega$ since:
\begin{equation}
\begin{array}{ll}
Pr(\nu | I) = \int_0^{2\uppi} {d\omega \; Pr(\nu,\omega | I) }\vspace{4pt} \\
\hspace{34pt}= \int_0^{2\uppi} {d\omega \; Pr(\nu | \omega, I) Pr(\omega | I) },
\end{array}
\end{equation}
and the speed of the planet changes as it orbits so that the probability density of finding the planet at a~true anomaly of $\nu$ varies with $\nu$ in an elliptical orbit: $Pr( \nu | \omega, I) \neq Pr( \nu' | \omega, I)$ where $\nu' \neq \nu$.

Consider a change in the viewing angle by $\phi$, so that $\nu \rightarrow \nu' = \nu + \phi$ and $\omega \rightarrow \omega' = \omega + \phi$.
It is then clear that:
\begin{equation}
Pr( \nu | \omega, I) = Pr( \nu' | \omega', I).
\end{equation}

Moreover, since:
$$
Pr( \omega | I ) = Pr( \omega' | I ) = \frac{1}{2\uppi} \; ,
$$
and since $Pr(\nu | \omega, I)$ is periodic, the two integrals over a single period:
\begin{equation}
Pr(\nu | I) = \int_{0}^{2\uppi} {d\omega \; Pr(\nu | \omega, I) \, Pr(\omega |I)}
\end{equation}
and:
\begin{equation}
Pr(\nu' | I) = \int_{\phi}^{2\uppi+\phi} {d\omega' \; Pr(\nu' | \omega', I) \, Pr(\omega' |I)}
\end{equation}
are equal so that:
\begin{equation}
Pr(\nu | I) = Pr(\nu' | I) = Pr(\nu + \phi | I)
\end{equation}
for arbitrary angle $\phi$. This implies that $Pr(\nu | I)$ is constant so that (\ref{eq:joint-angle-prior}) is generally true with:
\begin{equation}
Pr(\nu,i | I) = \frac{1}{4\uppi}
\end{equation}
for both circular and elliptical orbits.

The planet will be observed as transiting only for inclinations between $i_\text{min} \leq i \leq i_\text{max}$ where:
$$
i_\text{min} = \cos^{-1} \left( +\frac{R_{S}}{r} \right) \qquad \mbox{and} \qquad i_\text{max} = \cos^{-1} \left( -\frac{R_{S}}{r} \right),
$$
and $r$ is the star-planet distance. Therefore, the probability of a transit:
\begin{equation}
Pr(\mbox{transit} | I) = Pr(i_\text{min} \leq i \leq i_\text{max} | I),
\end{equation}
which is:
\begin{align}
Pr(i_\text{min} \leq i \leq i_\text{max} | I) &= \int_{i_\text{min}}^{i_\text{max}} {di \;\; \sin{(i)} \;\; Pr( i | I )} \nonumber \\
&= \int_0^{2\uppi} {d\nu \; \int_{i_\text{min}}^{i_\text{max}} {di \;\; \sin{(i)} \;\; Pr( \nu,i | I )}} \nonumber \\
&= \int_0^{2\uppi} {d\nu \; \int_{i_\text{min}}^{i_\text{max}} {di \;\; \frac{1}{4\pi} \; \sin{(i)}}} \nonumber \\
&= \frac{1}{4\uppi} \int_0^{2\uppi} {d\nu \; \left( \cos{(i_\text{max})} - \cos{(i_\text{min})} \right) } \nonumber \\
&= \frac{2}{4\uppi} \int_0^{2\uppi} {d\nu \; \frac{R_{S}}{r} }. \nonumber
\end{align}

Now, using the fact that the star-planet distance $r$ varies as a function of the true anomaly $\nu$ as described in (\ref{eq:eccentric-orbits}), the equation above becomes:
\begin{align}
Pr(i_\text{min} \leq i \leq i_\text{max}) &= \frac{1}{2\uppi} \int_0^{2\uppi} {d\nu \; \frac{R_{S} (1+e\cos{(\nu)})}{a (1-e^2)} }. \nonumber \\
&= \frac{R_{S}}{a (1-e^2)} \; , \nonumber
\end{align}
so that in the case of an eccentric orbit, the probability of observing the transit of a planet for which the eccentricity $e < 1$ is:
\begin{equation}
Pr(\mbox{transit}) = \frac{R_{S}}{a (1-e^2)} \; ;
\end{equation}
whereas for a circular orbit, or an orbit for which the eccentricity is zero, $e = 0$, the transit probability~is:
\begin{equation} \label{eq:transit-probability}
Pr(\mbox{transit}) = \frac{R_{S}}{a}.
\end{equation}

One can now compute the probability that a distant observer would view a planet in a circular orbit as transiting its host star. Table \ref{tab:transit-probability} lists these probabilities for several planets in our solar system for which the Sun has a radius of $R_{S} = 0.0046$ AU. According to (\ref{eq:transit-probability}), these probabilities fall off as $1/a$, indicating that the transit method of exoplanet detection is heavily biased toward the detection of closely-orbiting exoplanets.

\begin{table}[t]
\centering
\caption{This table lists the probability that a distant observer could possibly observe each of these planets transiting the Sun given the semi-major axis of their orbit in astronomical units (AU). These probabilities fall off as $1/a$ as expressed in (\ref{eq:transit-probability}).}
\label{tab:transit-probability}
\begin{tabular}{ccc}
\toprule
\textbf{Planet }& \textbf{Semi-Major Axis (AU)} & \textbf{Transit Probability} \\
\midrule
Mercury& 0.39 & 1.18\% \\

Venus & 0.72 & 0.64\% \\

Earth & 1.00 & 0.46\% \\

Mars  & 1.53 & 0.30\% \\

Jupiter & 5.20 & 0.09\% \\

Saturn & 9.54 & 0.05\% \\
\midrule

Solar Radius & 0.0046 AU & \\
\bottomrule
\end{tabular}
\end{table}

The probability of an alien civilization being able to observe Earth transiting the Sun (within the {{Earth transit zone}} (ETZ)) is only 0.46\%. Since the average stellar density in the galactic neighborhood surrounding the Sun is approximately 0.004 stars per cubic light year \cite{Gregersen:2010}, there are about 17,000 stars within 100 LY of the Sun. Of these 17,000 stars, only about 78 of them are positioned so that Earth could be observed transiting the Sun. A recent study utilized the Hipparcus database to identify 37 K and 45 G dwarf stars in the ETZ within 1000 parsecs ($\approx$3261 LY) \cite{Heller+Pudritz:2016}.

The {{secondary eclipse}} is similar to the transit. It occurs when the planet passes behind the host star so that the planetary flux, which is composed of both reflected light and thermal emissions, is~blocked. The magnitude of the secondary eclipse is dependent on the magnitude of the planetary flux, which, for a relatively cool planet, is bounded by the maximum amount of reflected light (\ref{eq:reflected-light}):
\begin{equation}
F_\text{secondary} < A_g\frac{{R_p}^2}{{a}^2} F_{S} = A_g\frac{{R_S}^2}{{a}^2} F_\text{primary},
\end{equation}
where we are considering a circular orbit of radius $a$, $A_g$ is the geometric albedo, $R_p$ is the radius of the planet, $R_S$ is the radius of the host star and $F_{S}$ is the stellar flux. Since typically $R_S \ll a$, secondary eclipses are not always detectable in the recorded data. EXONEST models secondary eclipses using the methodology developed by Mandel and Agol \cite{Mandel+Agol:2002} applied to a uniform source.  An example of a secondary eclipse can be seen in Figure \ref{fig:HAT-P_7b} at the orbital phase of 1.1 days.

\subsection{Reflected Light}

Planets reflect light both from cloud tops and from solid or liquid surfaces, and the degree to which they reflect light is quantified by their albedo \cite{Jenkins+Doyle:2003, Seager:2010, Perryman:2011, Placek:thesis:2014, Placek+Knuth+Angerhausen:EXONEST}. EXONEST employs the simplest model for reflected light in which the planet is assumed to be diffusely isotropically reflecting, so that the brightness does not depend on the angle of observation. In addition, it is assumed that the planet is situated sufficiently distant from the host star that the starlight impinges on the planet in parallel rays so that half of the planet is in daylight and the other half of the planet is in night.

Given these assumptions, the stellar flux reflected by the planet can be written as \cite{Sobolev:1975, Carter:thesis:2017}:
\begin{equation} \label{eq:reflected-light}
\Phi_{ref} = A_g \frac{{R_p}^2}{r^2} \left( \frac{(\uppi - \theta(t))\cos(\theta(t))+\sin(\theta(t))}{\uppi} \right),
\end{equation}
where $R_p$ is the radius of the planet, $A_g$ is the geometric albedo and $\theta(t)$ is the time-varying orbital phase angle.
This result appears to work well for many exoplanets.

\subsubsection*{New Efforts to Properly Model Reflected Light}

However, it is known that a significant number of planets (mainly hot Jupiters) closely orbit giant stars so that the assumption that the starlight impinges on the planet in the form of parallel rays is false. Instead, the planet can be divided into three zones. The first zone is that of full daylight where the entire apparent disk of the star can be seen from the surface. The second zone is a penumbral zone where only a portion of the star is visible as it appears to be setting or rising. The third zone is the night zone where the star is not visible. There are a number of cases in which the penumbral zone extends well into what would normally be the night zone. For example, as much as 70\% of {Kepler}-91b's surface is lit by starlight. This situation is illustrated in Figure \ref{fig:illuminated-planets}B. This has significant implications for the primary transit since a fraction of the backside of the planet may be partially illuminated in the penumbral zone, thus compensating somewhat for the reduced starlight due to the transit. As a result, unless the illumination geometry is properly modeled, primary transit depths will be less extensive resulting in an under-estimation of the planetary radius or an over-estimation of the nightside thermal flux followed by an over-estimation of nightside temperatures.

We are currently working to develop reflected light models for EXONEST that can accommodate the case in which planets are illuminated by parallel rays and the case in which planets are illuminated while closely-orbiting giant stars \cite{Kopal:1953, Kopal:1959, Carter:thesis:2017}. The fractional surface areas in the fully-illuminated zone, partially-illuminated penumbral zone and night zone are given by \cite{Carter:thesis:2017}:

\begin{align}
\sigma_\text{full} &= \frac{1}{2}\left(1-\frac{R_{S} + R_p}{r}\right) \label{eq:day}\\
\sigma_\text{pen} &= \frac{R_{S}}{r} \label{eq:pen}\\
\sigma_\text{night} &= \frac{1}{2}\left(1-\frac{R_{S} - R_p}{r}\right). \label{eq:night}
\end{align}

Since these surface areas depend on the star-planet separation distance $r$, eccentric orbits will require the use of the more general reflectance model.

\begin{figure}[t]
  \centering
  \includegraphics[width=0.8\textwidth]{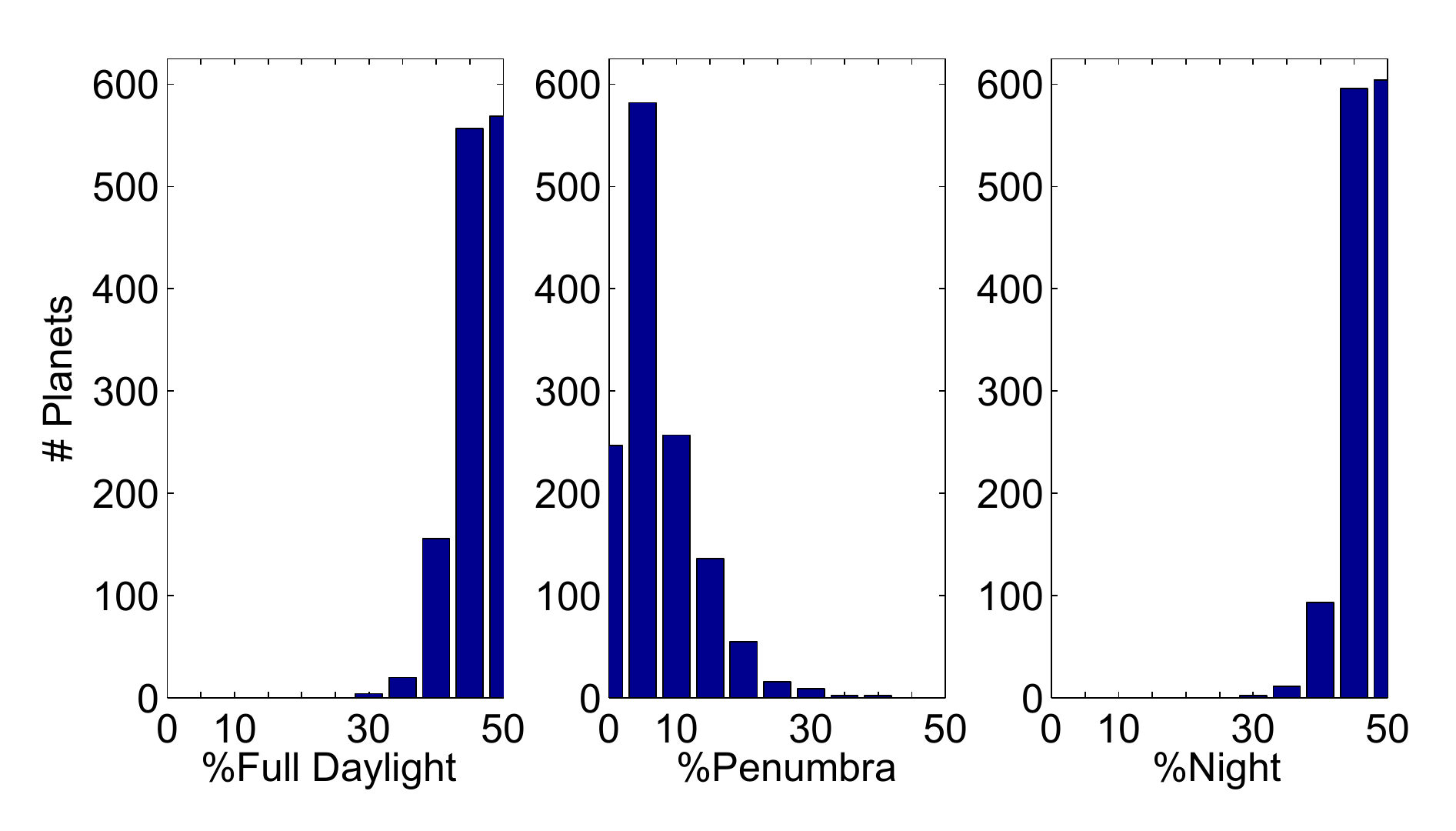}
  \caption{For 1306 exoplanets in the Exoplanet Archive, this figure shows histograms of the percent of the surface area that is in full daylight, in a penumbral zone and in completely dark night. Note that slightly less than half of the planets have a 50\% full daylight zone or a 50\% night zone. It is most common for planets to have a 5\% penumbral zone, and more than one quarter of the planets have penumbral zones that are larger than 5\%. There are four planets with significant penumbral zones in which the full daylight zone covers only about 30\% of the surface of the planet.}
  \label{fig:exoplanet-illumination}
\end{figure}

A quick survey of the exoplanet archive \cite{ExoplanetArchive:2017} reveals 1306 exoplanets for which the radius of the star, $R_{s}$, the radius of the planet, $R_{p}$, and the semi-major axis, $a$, are reported. Figure \ref{fig:exoplanet-illumination} illustrates histograms for these 1306 exoplanets revealing the percent of the surface area that is in full daylight, in a penumbral zone and in completely dark night. It is most common for planets to have a 5\% penumbral zone, and more than one quarter of the planets have penumbral zones that are larger than 5\%. There~are nine planets for which the penumbral zone covers about 30\% of the planet's surface and two planets for which the penumbral zone covers 40\% of the planet's surface!  Of these 1306~exoplanets, 39 of them have more than 60\% of their surface illuminated (dayside zone plus penumbral zone) by their host star. Another survey of the 1073 confirmed Kepler exoplanets that have orbital periods shorter than 10 days indicates that 88 of those planets (about 8\% of them) have more than 60\% of their surface illuminated (dayside zone plus penumbral zone) by their host star.
The significance of this effect in the context of the reflected light and the analysis of Kepler data will be explored in a future~paper.

We believe that improperly modeling stellar illumination was a problem in our earlier study involving a proposed Trojan partner to the planet {Kepler}-91b.~In that analysis, the Bayesian evidence weighed heavily in favor of a Trojan partner, but the modeled planetary temperatures were unphysically high, which led us to discount the possibility of a Trojan partner \cite{Placek+etal:K91:2015}. The star {Kepler}-91 is an asymptotic giant branch star with a radius of approximately $6.30 \pm 0.16 \, R_{\odot}$ (where $R_{\odot}$ is the solar radius), and the semi-major axis of the hot Jupiter {Kepler}-91b is $a = 0.072 \substack{+0.007 \\ -0.002}$ AU \cite{Esteves+etal:2015:K91}. The planet {Kepler}-91b has a radius of $R_p = 1.322 \substack{+0.094 \\ -0.086} \, R_J$, where $R_J$ represents Jupiter's radius \cite{Esteves+etal:2015:K91}. By~Equations (\ref{eq:day})--(\ref{eq:night}), this implies that $29.5\% \pm 1.0\%$ of {Kepler}-91b is in full daylight, $40.2\% \pm 2.1\%$ is in partial penumbral light and $30.4\% \pm 1.0\%$ is in full darkness, so that approximately $69.6\%$ of the planet is illuminated at any point in time (as is illustrated in Figure \ref{fig:illuminated-planets}B). By not properly accounting for this illumination, the modeled thermal emissions would be forced to compensate with increased day-side and night-side temperatures possibly explaining the unphysically high temperatures obtained in the analysis \cite{Placek+etal:K91:2015}, which led to the Trojan hypothesis being discounted. This is supported by the fact that other studies of this system have led to unphysically high temperatures. For example, a study of KOI-2133b ({Kepler}-91b) by Esteves et al. \cite{Esteves+DeMooij+Jayawardhana:2013} found that the nightside temperature was $3100 \pm 200$~K, which was greater than the expected equilibrium temperature of $1570$~K, leading them to hypothesize that {Kepler}-91b, which was later re-confirmed as a planet \cite{Sliski+Kipping:2014}, was self-luminous.

\subsection{Thermal Emissions}

The modeling of thermal emissions is similar to that of reflected light. Current models assume that there is a day-side and a night-side and that these two sides are characterized by two
effective
temperatures $T_{D}$ and $T_{N}$, respectively.

The planet is assumed to radiate as a thermal blackbody so that the detected thermal flux is given by \cite{Charbonneau+etal:2005}:
\begin{equation} \label{eq:stellar-flux}
F = \int{B(\lambda; T) K(\lambda) \; d\lambda}
\end{equation}
where $T$ is the effective temperature of the radiating region, $B(\lambda;T)$ represents the spectral radiance of a blackbody at temperature $T$:
\begin{equation}
B(\lambda;T) = \frac{2hc^2}{\lambda^5}\frac{1}{\exp\left({\frac{hc}{\lambda k_B T}}\right)-1} \; ,
\end{equation}
$h$ is Planck's constant, $k_B$ is Boltzmann's constant, and $K(\lambda)$ is the Kepler response as a function of wavelength $\lambda$ \cite{KeplerHandbook}.

The relative day-side and night-side fluxes are given by:
\begin{align}
\frac{F_{D}(t)}{F_S} &= \frac{1}{2} (1 + \cos({\theta(t)})) \left(\frac{R_p}{R_S}\right)^2 \frac{\int{B(\lambda; T_{D}) K(\lambda) \; d\lambda}}{\int{B(\lambda; T_{S}) K(\lambda) \; d\lambda}} \label{eq:thermal-day}\\
\frac{F_{N}(t)}{F_S} &= \frac{1}{2} (1 + \cos({\theta(t)-\uppi})) \left(\frac{R_p}{R_S}\right)^2 \frac{\int{B(\lambda; T_{N}) K(\lambda) \; d\lambda}}{\int{B(\lambda; T_{S}) K(\lambda) \; d\lambda}} \; ,
\end{align}
with a total relative flux of:
\begin{equation}
\frac{F_\text{thermal}(t)}{F_S} = \frac{F_{D}(t)}{F_S} + \frac{F_{N}(t)}{F_S}.
\end{equation}

The similarity of the thermal emission phase curve to the reflected light phase curve means that for a circular orbit, the models are degenerate so that it is not possible, using single-bandpass photometry, to distinguish thermal emission from reflected light. Only for significantly eccentric orbits, with $e > 0.3$, in which the planet-star distance $r(t)$ changes, thus affecting reflected light, can thermal flux be modeled independently of reflected light \cite{Placek+Knuth+Angerhausen:EXONEST}, assuming that the temperatures on the planet are relatively constant. Observations in multiple bandpasses, such as combining data from both Kepler and TESS, will enable thermal emissions to be distinguished from reflected light \cite{Placek+Knuth+Angerhausen:2016:Kepler+TESS}. Just as in the case of reflected light, a closely-orbiting planet will have three zones: a dayside zone, a penumbral zone and a nightside zone. While one could model each zone with a separate average temperature, previous~studies have found that the contribution of the penumbral zone to thermal emissions is negligible \cite{Leger+etal:2011, Maurin+etal:2012}.

\subsection{New Efforts to Model Refracted and Forward Scattered Light}

\begin{figure}[t]
  \centering
  \includegraphics[width=0.72\textwidth]{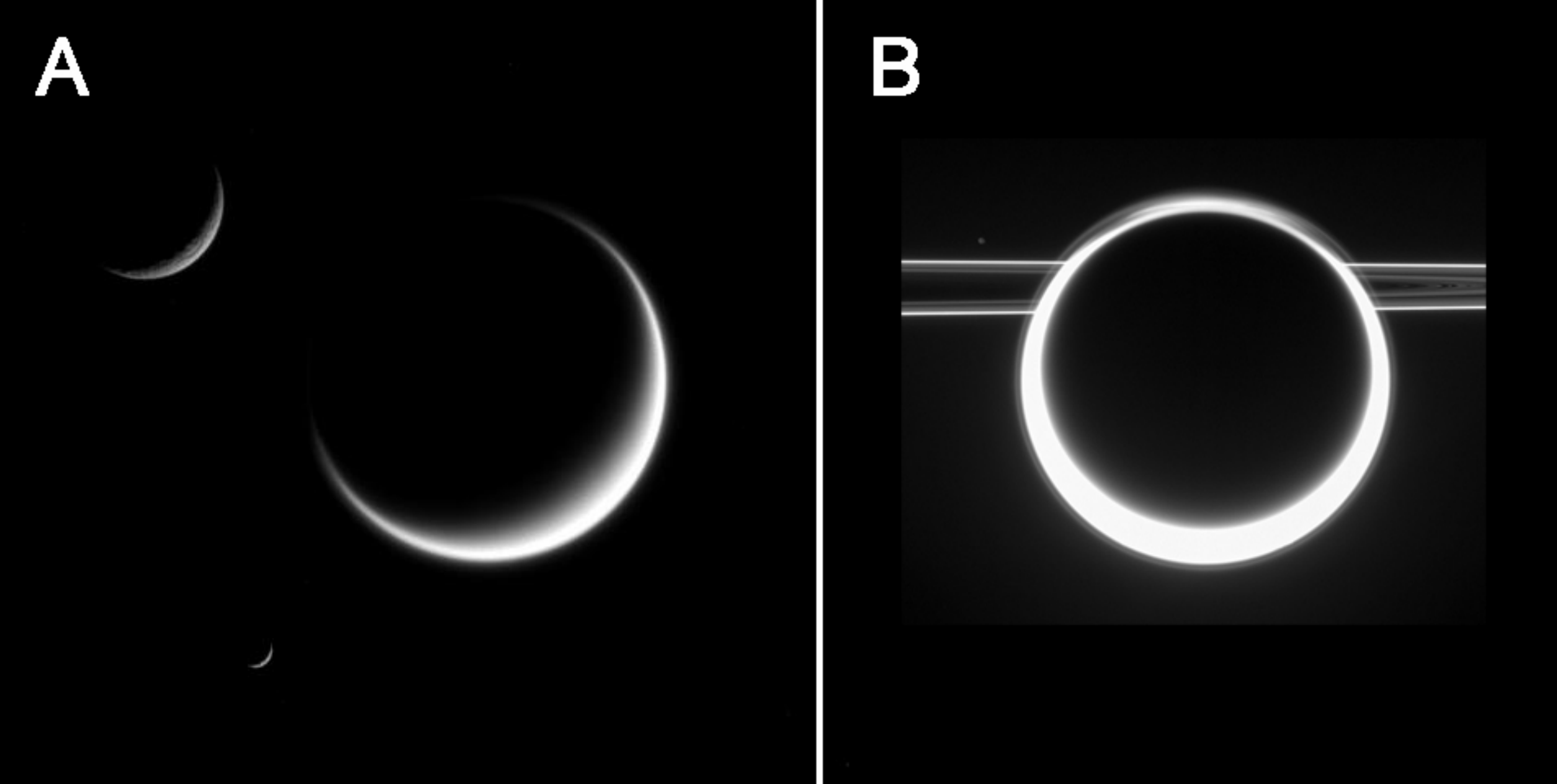}
  \caption{\textbf{A} This photograph taken by the Cassini space probe of Saturn’s moons Rhea (upper left), Titan (center) and Mimas (lower left) illustrates the role that Titan’s atmosphere plays in changing the illumination of the moon when almost backlit. The illuminated arc is much brighter and more extensive on Titan due to forward scattering through its significant atmosphere. \textbf{B} In the image on the right, Titan is seen in the foreground with Saturn’s rings in the background. Both are backlit in this image, and Titan is surrounded by a brilliant halo of forward-scattered light. Public Domain Images. Credit: NASA/JPL-Caltech/Space Science Institute.} \label{fig:titan+rhea+mimas}
\end{figure}

The situation is significantly more complicated for a planet or moon that possesses an atmosphere. As the planet approaches the primary transit, it becomes more and more backlit. In addition to reflected light from atmospheric clouds or the surface, one may potentially detect light that is both forward scattered by the atmosphere and refracted through the atmosphere \cite{Sidis:2010, Munoz+etal:2012, Misra+Meadows+Crisp:2014, Misra:2014, Betremieux+Kaltenegger:2015, Munoz+Cabrera:2017}. One of these effects is illustrated in Figure \ref{fig:titan+rhea+mimas} where in Panel (A), we see an image taken by the Cassini probe of Saturn's moons Rhea (upper left), Titan (center right) and Mimas (lower center). Both Rhea and Mimas are airless worlds, and one sees only reflected light from their surfaces. Titan, on the other hand, is a large moon with an extended nitrogen and methane atmosphere and, more critically, rich in strongly forward scattering haze. Here, one sees that the starlight entering the atmosphere is forward scattered by the haze particles, thereby forming an illuminated arc much brighter and more extensive than on the other two moons for which such a ring is not possible \cite{Munoz+Lavvas+West:2017:Titan}. Forward scattered by hazy exoplanet atmospheres, if occurring, will be more important for close-in planets, whereas the refraction signal is likely more important for far-out transiting planets and immediately before/after transit, where refraction in the upper atmosphere may cause shoulders in the pre-/post-transit light curves \cite{Munoz+etal:2012, Misra+Meadows+Crisp:2014, Misra:2014, Dalba:2017}.

The increase in photometric flux due to refraction is on the order of \cite{Sidis:2010}:
\begin{equation} \label{eq:refraction}
\frac{F_\text{refraction}}{F_{S}} \propto \frac{H R_{p}}{{R_{S}}^2} \; ,
\end{equation}
where $H$ is the scale height of the planetary atmosphere, which is given by:
\begin{equation}
H = \frac{k_{B} T}{\mu g} \; ,
\end{equation}
where $k_{B}$ is Boltzmann's constant, $T$ is the average atmospheric temperature, $\mu$ is the mean molecular weight of the atmospheric constituents
and $g$ is the gravitational acceleration at the surface of the planet \cite{Sidis:2010}. Typically, $H \ll R_{p}$, so that by (\ref{eq:refraction}), the refractive effects are much smaller than the primary transit depth:
\begin{equation}
\frac{F_\text{refraction}}{F_{S}} \ll \frac{{R_{p}}^2}{{R_{S}}^2}.
\end{equation}

For example, in the case of the Earth ($T = 288$ K, $g = 9.8$ m s$^{-2}$, $\mu = 0.0288$ g mol$^{-1}$), the~characteristic atmospheric height is estimated to be $H = 8.48\times10^3$ m, which is reasonable considering that the troposphere has an average height of $12\times10^3$ m. As a result, the photometric flux of light refracting through Earth's atmosphere at the point of occultation is about $\frac{H}{R_{p}} \approx \frac{8.48\times10^3\,\text{m}}{6.37\times10^6\,\text{m}} = 0.0013$ times, or three orders of magnitude, smaller than the primary transit depth.

In the event that the photometric effects of refracted light can be detected, this would provide information about the nature of the exoplanetary atmosphere.

\subsection{New Efforts to Model Atmospheric Effects}

Atmospheric effects can play a significant role in photometry. For example, atmospheric clouds can shift a planet's highly reflective bright spot away from the substellar point. Similarly, atmospheric {{superrotation}} can shift a planet's thermally bright hotspot away from the substellar point. Such effects, which can be modeled as an angular shift in the reflectance or thermal phase curves, can reveal a great deal about the exoplanetary atmosphere \cite{Showman+Polvani:2011:superrotation, Esteves+etal:2015:K91, Faigler+Mazeh:2015, Placek+Angerhausen+Knuth:2017:optimization}. More exotic effects are possible. For example, it is thought that supersonic flow and shocks can develop on the dayside of irradiated hot Jupiters upstream of the substellar point and that these shocks will affect the position of the peak of the thermal phase curve \cite{Heng:2012:shocks}. Heavily-irradiated hot Jupiters are expected to have an electrically-conductive atmosphere due to ionization. As a result, exoplanetary magnetic fields are expected to affect atmospheric flow~\cite{Batygin+Stanley+Stevenson:2013} and in some cases, magnetically-dragged winds are expected to reduce the longitudinal offset of the thermal hotspot \cite{Menou:2012, Batygin+Stanley+Stevenson:2013, Rauscher+Menou:2013}.

Other deviations in the reflectance and thermal phase curves may reveal more about planetary atmospheres, such as inhomogeneous cloud cover across the planet's surface \cite{Demory+etal:2013:clouds, Heng+Demory:2013}, atmospheric variability \cite{Armstrong+etal:2016}, evolving weather patterns \cite{Palle+etal:2008, Artigau+etal:2009, Gillon+etal:2013}, diurnal atmospheric effects \cite{Ford+Seager+Turner:2001} or seasonal atmospheric effects \cite{VanEylen+etal:2013}.

\subsection{Doppler Boosting/Beaming}

The gravitational influence of a planet causes both the star and the planet(s) to orbit a common center of mass. As a result, stars with planets tend to wobble. This is the basis of the {{radial velocity technique}} in which the radial velocity of the star is carefully measured using spectroscopy and knowledge about the Doppler shift to detect the presence of planets. While the radial velocity technique relies on spectroscopy, there is a photometric component that arises due to relativistic considerations related to stellar aberration in which an object that radiates light uniformly at rest will, in motion, tend to radiate more light in the direction of motion and less light opposite that direction \cite{Rybicki+Lightman:2008}. This effect is called {{Doppler boosting}} or {{Doppler beaming}}.

The effect can be derived using special relativity resulting in the following formula for the boosted flux \cite{Rybicki+Lightman:2008, Placek:thesis:2014, Placek+Knuth+Angerhausen:EXONEST}:
\begin{equation}
F_\text{boost}(t) = F_{S} \left( \frac{1}{\gamma (1-\beta \cos{\theta(t)})} \right)^4
\end{equation}
where $\gamma^{-1} = \sqrt{1-\beta^2}$, and $\beta = \frac{v}{c}$ where $c$ is the speed of light and $F_{S}$ is the stellar flux in the reference frame of the star. Since the motion of the star will involve velocities no more than 100--1000 m/s ($\beta \approx 10^{-5}$), it is perhaps surprising that this relativistic effect could be detectable. However, the fact that the star is approximately $10^6$-times brighter than the reflected light from a planet means that this effect can be on the order of that of reflected light. Given these speeds, a non-relativistic approximation is reasonable:
\begin{equation}
\frac{F_\text{boost}}{F_{S}} \approx 1 + 4\beta_{r}(t)
\end{equation}
where $\beta_{r}$ represents the radial component of the stellar velocity as viewed from Earth scaled by the speed of light so that:
\begin{equation}
\beta_{r}(t) = \frac{V_{z}(t)}{c}
\end{equation}
where $V_{z}$ is the radial velocity of the planet from the perspective of Earth:
\begin{equation}
V_{z}(t) = K \left( \cos{(\nu(t)+\omega)} + e \cos{\omega} \right)
\end{equation}
 in which $\nu(t)$ is the true anomaly as a function of time, $\omega$ is the argument of the periastron, $e$ is the eccentricity (see Figure \ref{eq:eccentric-orbits})
and $K$ is the radial velocity semi-amplitude \cite{Placek:thesis:2014} given by:

\begin{equation}
K = \left( \frac{2 \uppi G}{T} \right)^{-\frac{1}{3}} \frac{ M_{p} \sin{(i)} }{ {M_{S}}^{\frac{2}{3}} \sqrt{1-e^2}},
\end{equation}
 in which $G$ is Newton's gravitational constant, $T$ is the period of the planet, $M_{p}$ is the mass of the planet, $i$ is the inclination of the orbital plane with respect to Earth and $M_S$ is the mass of the star.
Note~that when the inclination of the orbital plane is zero, $i = 0$, the orbit is face on (in the projected plane of the sky), and both the radial velocity semi-amplitude and the radial velocity are zero, resulting in no Doppler effect.
Furthermore, as the planet orbits the star, the true anomaly $\nu(t)$, which is an angle, advances. There will be two times per orbital period at which the trigonometric factor goes
to zero, resulting in zero radial velocity and therefore zero Doppler effect.
This is most easily seen by considering a circular orbit for which the eccentricity is zero, $e = 0$, and the argument of the periastron is zero, $\omega = 0$. In the case of this circular orbit, the radial velocity is zero when the true anomaly is $\nu = \frac{\uppi}{2}$ and $\nu = \frac{3\uppi}{2}$, which occurs when the planet is moving in the projected plane of the sky.
Since the Doppler boosting effect involves both the mass of the planet and the inclination angle of the orbital plane, this~phenomenon facilitates the estimation of these two quantities.

\subsection{Tidal Forces and Ellipsoidal Variations}
\label{ModelsIntro}

Tidal interactions between a massive object and a star create distortions on the star  (see Figure \ref{fig:evil-mc-star}),
which are potentially observable by the Kepler Space Telescope. Known as {{ellipsoidal variations}}, this photometric effect appears with two maxima per orbit, each occurring when the largest cross-sectional area of stellar surface is observed. Bayesian model testing provides a rigorous framework to compare the various representations of ellipsoidal variations. A preferred representation of ellipsoidal variations will become increasingly important for exoplanet classification with higher precision photometry.

\subsubsection{Trigonometric Models}

The BEaming, Ellipsoidal and Reflection (BEER) model, which was developed by Faigler and Mazeh \cite{Faigler+Mazeh:2007}, models the ellipsoidal variations as being proportional to the cosine of twice the phase~angle:
\begin{equation}
\frac{F_\text{ellip}(t)}{F_S} \approx -\beta \frac{M_p}{M_S} \frac{{R_{S}}^3}{a^3} \sin^2(i)\cos\Big(2\theta(t)\Big) \; ,
\end{equation}
in which on the left-hand side, the flux due to the ellipsoidal variations $F_\text{ellip}$ is normalized by the stellar flux $F_S$, $M_p$ is the planet's mass, $M_S$ is the star's mass, $R_S$ is the star's radius, $a$ is the star-planet distance, $i$ is the inclination of the orbital plane and $\theta$ is the orbital phase [0, $2\uppi$]. The BEER model allows for positive and negative values for the amount of ellipsoidal variation observed.
The effects of linear gravity-darkening and limb-darkening,
 $\beta$, are modeled by \cite{Morris:1985}:
\begin{equation}
\beta = 0.15 ~\frac{(15+u)(1+g)}{(3-u)} \; ,
\end{equation}
in which the linear limb-darkening and gravity-darkening coefficients are represented by $u$ and $g$, respectively. Estimates of the coefficients $u$ and $g$ are provided by modeling metallicity and effective stellar temperature \cite{Heyrovsky:2007, Sing:2010}.

The Kane and Gelino model hypothesizes ellipsoidal variations to be proportional to the projected separation distance between the star and planet \cite{Kane+Gelino:2012},
which, when substituted for the orbital phase,~become:
\begin{equation}
\frac{F_\text{ellip}(t)}{F_{S}} \approx \beta \frac{M_p}{M_S} \frac{{R_{S}}^3}{a^3} \sin\big(\theta(t)\big).
\end{equation}

However, this model suffers from a discontinuity in the first derivative, which is not seen in simulations of ellipsoidal variations using gravitational isopotentials.

A modified form, labeled Kane and Gelino (Mod), was introduced by Placek et al. \cite{Placek+Knuth+Angerhausen:EXONEST}, which~removes the discontinuity by setting the effect proportional to the square of the projected separation distance:
\begin{equation}
\frac{F_\text{ellip}(t)}{F_S} \approx \beta \frac{M_p}{M_S} \frac{{R_{S}}^3}{a^3} \sin^2\big(\theta(t)\big).
\end{equation}

Unlike the BEER model, which produces negative values, the two Kane and Gelino models have a~minimum value of zero.

\subsubsection{Direct Modeling: EVIL-MC}

\begin{figure}[t]
  \centering
  \includegraphics[width=0.5\textwidth]{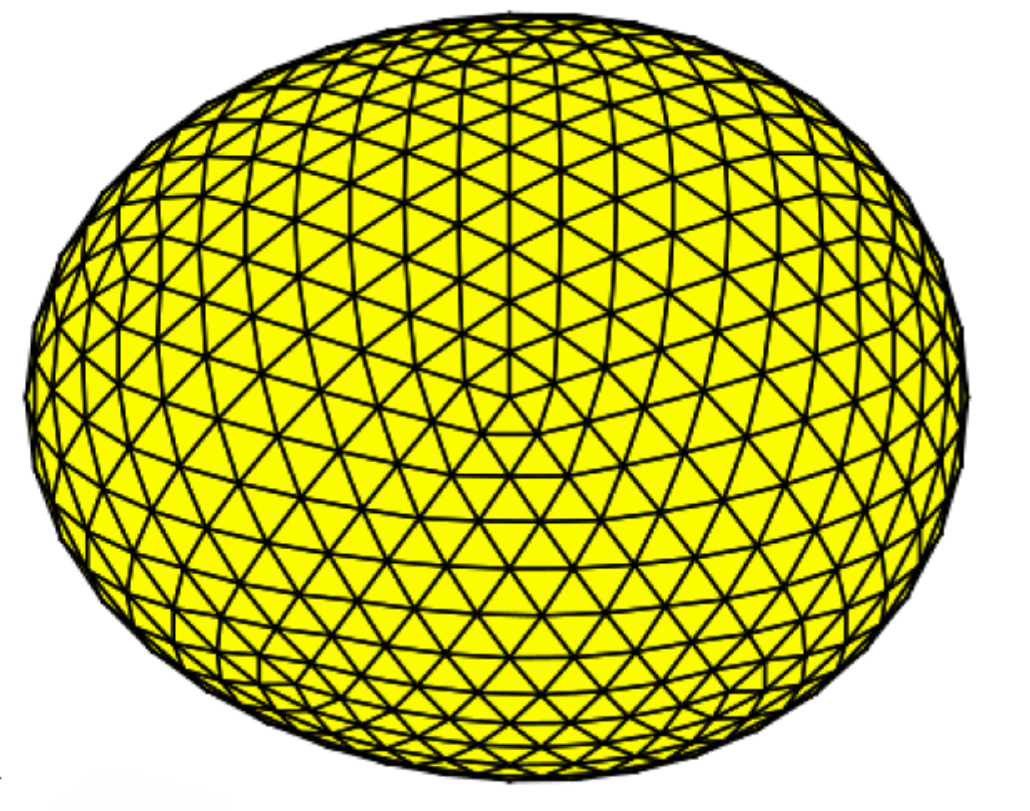}
  \caption{This is an exaggerated image of a star that has been tidally distorted by a closely-orbiting planet $\left(\frac{M_p}{M_S} = 9.5425\right)$. The increased cross-sectional area results in variations in flux that oscillate at twice the orbital period as the planet revolves around the star.}
  \label{fig:evil-mc-star}
\end{figure}

The Ellipsoidal Variations Induced by a Low-Mass Companion (EVIL-MC) model was developed by Jackson et al. \cite{Jackson+etal:2012} to directly model tidal effects on the stellar atmosphere. EVIL-MC was modified for use in EXONEST \cite{Gai:thesis:2016, Gai+Knuth:2017} in which a geodesic structure of 640~triangles is used to represent the stellar surface as illustrated in Figure \ref{fig:evil-mc-star}.
The deviations from sphericity are computed for each point on the surface model. The observed flux, $\phi_\text{ellip}$, is computed by summing the blackbody flux within the Kepler bandpass (\ref{eq:stellar-flux}) over the observed surface area. The ellipsoidal variation effect is then computed~with:
\begin{equation}\label{PhiEllipse_EVILMC}
\frac{F_\text{ellip}}{F_{S}} = \frac{\phi_\text{ellip} - \phi_\text{sphere}}{\phi_\text{sphere}} \; ,
\end{equation}
where $\phi_\text{sphere}$ is the flux from a spherical star and $\phi_\text{ellip}$ is the flux from the ellipsoidal star. While direct modeling should provide a more accurate representation, EVIL-MC is significantly ($\approx$1000--10,000 times) more computationally expensive than the trigonometric models. The Kane and Gelino (Mod) model provides a computationally fast approximation to EVIL-MC.

\subsubsection{Model Testing }

We have recently performed model testing on Kepler light curve data using EXONEST \cite{Gai:thesis:2016, Gai+Knuth:2017}. EVIL-MC provided the best representation of ellipsoidal variations in the Kepler-13 system followed by the Kane and Gelino (Mod) and BEER models, which had similar evidence values. The values are reported in Table \ref{tab:Table_Evidence}. The Kane and Gelino (Mod) model is closest to the EVIL-MC results and has slightly higher evidence over the others suggesting that, given its straight-forward analytic form, the~Kane and Gelino (Mod) is the most effective model for this effect.

\begin{table}[t]
\begin{center}
\caption{This table lists the evidence values obtained by performing model testing with the EXONEST algorithm. The Bayes' factor is computed by taking the difference between the logarithm of the Bayesian evidence of the model under consideration and the Ellipsoidal Variations Induced by a Low-Mass Companion (EVIL-MC) model. EVIL-MC is slightly preferred over the trigonometric models of Kane and Gelino (Mod) and BEaming, Ellipsoidal and Reflection (BEER).}

\begin{tabular}{ccc}
\toprule
\textbf{Model }& \textbf{Log (Evidence)} & \textbf{EVIL-MC's Bayes' Factor} \\ \midrule
EVIL-MC & 14,256.91 $\pm$ 0.91 & 1 \\
Kane and Gelino (Mod) & 14,256.71 $\pm$ 0.73 & 2.20 \\
BEER & 14,256.26 $\pm$ 0.74 & 2.65 \\
Kane and Gelino (2012) & 14,253.66 $\pm$ 0.72 & 5.25 \\ \bottomrule
 \end{tabular}\label{tab:Table_Evidence}
\end{center}
\end{table}
\unskip

\section{Bayesian Inference Engine}

For a given exoplanetary model, the Bayesian inference engine allows us to compute the Bayesian evidence and the model parameter values along with their associated uncertainties. The current implementation allows for a choice of sampling methods: Metropolis--Hastings Markov chain Monte Carlo (MCMC) \cite{Metropolis:1953, Hastings:1970} (which does not provide the Bayesian evidence), the original nested sampling algorithm \cite{Skilling:2004:nested, Skilling:2006:nested, Sivia&Skilling} or the more recent variant called MultiNest \cite{Feroz+etal:2009, Feroz+etal:2011a, Feroz+etal:2013}. While the MultiNest algorithm has proven to shorten run times from several days to 10s of hours, it is usually not possible to operate with more than 75 samples, which severely reduces the precision to which the Bayesian evidence and the parameter values can be estimated.

Unfortunately, even by integrating a single period in cases with one exoplanet, the time required to evaluate the log likelihood is too great to allow for a rapid analysis. It may be of benefit to re-parameterize the eccentricity and argument of periapsis as suggested by Ford \cite{Ford:2006:MCMC}. In addition, since~we have found almost no evidence of phase changes in our analyses to date, it has been suggested (P.~Goggans, personal communication) that EXONEST be designed to utilize simulated annealing.

Model testing is a significant part of the EXONEST analysis pathway. By turning on-and-off various photometric models and comparing the Bayesian evidence in each case, one can build a case that a candidate exoplanet is a real planet and not a false positive \cite{Placek+Knuth+Angerhausen:EXONEST, Knuth+etal:DSP:2015}.

\subsection*{Priors and Likelihoods}

EXONEST allows priors on the parameter values to be specified. However, since we are still learning a great deal about exoplanets, the assignment of an informative prior based on known exoplanets is potentially detrimental as it may bias against the discovery of new phenomena. For this reason, uniform or non-informative priors are most often used.

Likelihoods can also be specified for EXONEST.
While Gaussian likelihoods are typically employed, there have been situations involving asymptotic giant branch stars in which likelihoods that accommodate correlated noise (red noise) are necessary \cite{Placek+etal:K91:2015}, prompting us to implement Sivia's nearest-neighbor approach \cite{Sivia&Skilling}:
\begin{equation}
\log L = -\frac{N}{2} \log{(2 \uppi \sigma^2)} - \frac{N-1}{2} \log{(1-\epsilon^2)} + \frac{Q}{2(1-\epsilon^2)}
\end{equation}
where $N$ is the number of data points, $\sigma^2$ is the noise variance, $\epsilon$ is the correlation strength, which is a parameter to be estimated, and $Q$ depends on the sum of the square of the residuals $\chi^2$ by:

\begin{equation}
Q = \chi^2 + \epsilon [ \epsilon (\chi^2 - \phi) - 2\psi ].
\end{equation}

The quantity $\phi$ is the sum of the first and the last squared residuals, and $\psi$ is the sum of the nearest neighbor squared residuals:
\begin{equation}
\psi = \sum_{i=1}^{N-1}{ R_i \, R_{i+1} }.
\end{equation}

As usual, the residuals are given by the difference between the modeled relative flux and the observed relative flux. In practice, one divides the observed flux by the average of the observed flux to obtain the observed relative flux. The residuals are typically given by:
\begin{equation}
R_i \equiv R(t_i) = \frac{\phi(t_i)}{\langle \phi(t_i) \rangle} - \frac{F_\text{total}(t_i)}{F_S} \; ,
\end{equation}
where $\phi(t_i)$ is the observed or measured photometric flux at time $t_i$ and $\frac{F_\text{total}(t_i)}{F_S}$ is the sum of the modeled relative photometric fluxes at time $t_i$. Unfortunately, this practice is not ideal since it is not generally true that $F_S \approx \langle \phi(t_i) \rangle$ since the photometric effects (transits, etc.) do not all integrate to zero. In principle, one could subtract the flux during the secondary eclipse, since ellipsoidal variations will be minimal and there will be no planetary effects. However, it would be better to model the stellar flux and to marginalize over it.

It is a straightforward problem to incorporate additional data, such as radial velocity measurements, by adding terms to the log likelihood. This has been done in our study of the {Kepler}-91 system \cite{Placek+etal:K91:2015}.

\section{Application: {Kepler}-76b}

{Kepler}-76b is a transiting hot-Jupiter with a radius of approximately $1.25~R_{\text{J}}$ that orbits a $1.2~M_{\text{sun}}$ star with an orbital period of $1.5449$~days \cite{Faigler+etal:2013}.~This planet is known to exhibit reflection/thermal emission, Doppler boosting, as well as ellipsoidal variations \citep{Faigler+etal:2013, Angerhausen+DeLarme+Morse:2015, Esteves+etal:2015:K91}.~This section outlines EXONEST's model selection capabilities with fits of four different forward models that each account for different combinations of photometric effects. The simplest model includes only the reflection component (R), which in the case of a circular orbit subsumes thermal emission. The second model includes reflection and ellipsoidal variations (RE). The third model accounts for reflection, ellipsoidal variations and Doppler beaming (REB). Finally, the last model accounts for each of the three effects in the REB model in addition to an angular shift in the phase curve (reflection component), which is indicative of atmospheric superrotation (REB-SR).  The data, which were obtained from the Mikulski Archive for Space Telescopes (MAST) managed by the Space Telescope Science Institute (STScI), consists of all available quarters of long-cadence Kepler Space Telescope data for {Kepler}-76b.  The~data were preprocessed to remove the primary transit so that only the overall phase curve and the secondary eclipse were fit by the EXONEST models. Note that even without considering the primary transit, excellent parameter estimates were obtained.

\begin{figure}[t]
\center
\includegraphics[scale = 0.58]{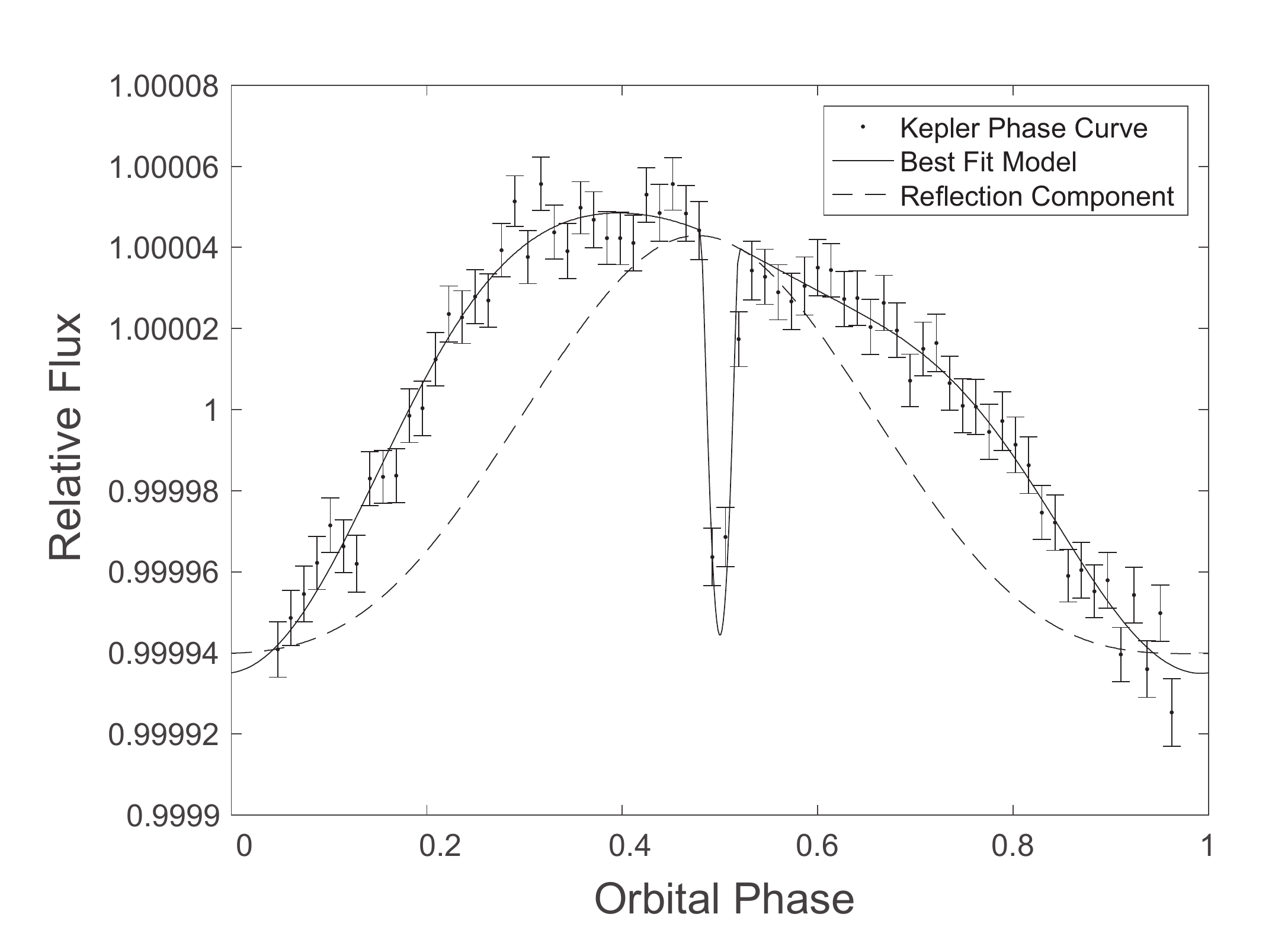}
\caption{An illustration of the best fit model (REB + SR) for {Kepler}-76b. The overall shape of the light curve is due to the reflected light, Doppler boosting and ellipsoidal variations. The secondary eclipse can be observed at an orbital phase of 0.5, which is shifted from the peak of the reflected light component (dashed curve) by 8.53${}^{\circ} \pm$ 0.12${}^{\circ}$, which most likely is due to atmospheric superrotation.}
\label{fig:Kepler-76b}
\end{figure}

The results of the model testing are listed in Table \ref{tab:kepler76_evidences}. In this example, the most complex model (REB-SR),  illustrated in Figure \ref{fig:Kepler-76b},  was favored by both the Bayesian evidence ($\log Z$) and the maximum log-likelihood ($\log L_{\text{max}}$) to explain the data. We can thus claim the detection of a shift in the phase curve by 8.53${}^\circ \pm$ 0.12${}^\circ$ degrees westward of the sub-stellar point, which is most likely caused by atmospheric superrotation. This result is consistent with previous studies of {Kepler}-76b, which yielded estimated westward shifts of 10.3${}^\circ \pm$ 2.0${}^\circ$ \cite{Faigler+etal:2013} and 8.28${}^\circ \pm$ 1.22${}^\circ$ \citep{Esteves+etal:2015:K91}. The parameter estimates for the REB-SR model are listed in Table~\ref{tab:kepler76_estimates}. Note that these estimates are comparable to those reported \cite{Faigler+etal:2013} when {Kepler}-76b was~discovered.

\begin{table}[t]
\begin{center}
\begin{tabular}{cc c }
\toprule
\textbf{Model }& \boldmath{$\log$} \textbf{Z} \textbf{(}\boldmath{$\times$}\textbf{10}\boldmath{$^6) $} & \boldmath{$\log$} $ L_{\text{\textbf{max}}} (\times 10^6)$ \\
\midrule
R & $-$9.5987 & $-$9.5987 \\
RE &$-$9.5822 & $-$9.5822 \\
REB &$-$9.5685 & $-$9.5684 \\ \midrule
\textbf{REB + SR} & \textbf{$-$9.5676} & \textbf{$-$9.5676} \\ \bottomrule
\end{tabular}
\end{center}
\caption{ EXONEST model testing results for {Kepler}-76b. This table lists the log-Bayesian evidence ($\log Z$) and maximum log-likelihood values ($\log L_{\text{max}}$) for model fits of {Kepler}-76 data. The favored model (listed in bold) incorporates each of the photometric effects plus atmospheric superrotation (reflection, ellipsoidal~variations and Doppler beaming (REB)-SR). This model is favored by both the Bayesian evidence and the maximum log-likelihood.} \label{tab:kepler76_evidences}
\end{table}
\unskip
\begin{table}[t]
\begin{center}
\begin{tabular}{ c c c }
\toprule
\multicolumn{3}{c}{{\textbf{Kepler}}\textbf{-76b Parameter Estimates (REB + SR Model)}} \\
\midrule
\textbf{Model Parameters} & \textbf{REB + SR Parameter Estimates} & \textbf{Literature Values}\\
\midrule
$R_p$ & 1.35 $\pm$ 0.05~$R_J$ & 1.25 $\pm$ 0.08~$R_J$$\;{}^{(a)}$\\
$M_p$ & 1.97 $\pm$ 0.12~$M_J$ & 2.0 $\pm$ 0.26~$M_J$$\;{}^{(a)}$\\
$i$ & 77.9${}^{\circ}$ $\pm$ 0.03${}^{\circ}$ & 78.0${}^{\circ}$ $\pm$ 0.2${}^{\circ}$$\;{}^{(a)}$\\
$A_g$ & 0.31 $\pm$ 0.22 & 0.23 $\pm$ 0.02$\;{}^{(a)}$, 0.22 $\pm$ 0.02$\;{}^{(b)}$\\ \bottomrule
\end{tabular}
\end{center}
\caption{ EXONEST (REB + SR model) parameter estimate results for {Kepler}-76b. This table lists the estimated radius, $R_p$, mass, $M_p$, orbital inclination, $i$, and geometric albedo, $A_g$, of {Kepler}-76b. These~values are comparable to those reported by $(a)$  Faigler et al. using the BEaming, Ellipsoidal and Reflection (BEER) algorithm \cite{Faigler+etal:2013}, and $(b)$  Angerhausen et al. \cite{Angerhausen+DeLarme+Morse:2015}.} \label{tab:kepler76_estimates}
\end{table}
\unskip
\vspace{-18pt}

\section{Conclusions}

EXONEST: The Exoplanetary Explorer is a software package for detecting and characterizing exoplanets from Kepler and CoRoT data. The package models a large number of photometric effects, each of which provides valuable information about the exoplanet and the exoplanetary system. New~efforts include developing more precise models of reflected light, refracted light and atmospheric effects, as well as re-coding to transform EXONEST into a Python-based open source project with the capability to employ third-party plug-and-play models of exoplanet-related photometric effects. In~addition, new planetary models are being developed that are relevant to rapidly rotating planets with shapes ranging from that of an oblate ellipsoid to pyriform shapes \cite{Abraham+Shaw:1983:Dynamics} and synestias \cite{Lock+Stewart:2017:synestias}. These~features are summarized in Table \ref{tab:effects}.

\begin{table}[t]
\begin{center}
\caption{ This table lists the modeled effects considered in this paper and indicates which effects are currently implemented (`X') in EXONEST, and which are currently under development (``D'') or being refined (``R'') or studied (``S'') for potential inclusion in the next generation of our software. Effects~marked with an ``M'' are being or will be evaluated using Bayesian model testing.}

 \begin{tabular}{c c c }
  \toprule
  \textbf{Modeled Effect} & \textbf{EXONEST} & \textbf{Under Study} \\ \midrule
  Circular Orbits & X & \\
  Elliptical Orbits & X & \\
  Primary Transit & X & \\
  Secondary Eclipse & X & \\
  Limb Darkening & X & M \\
  Reflected Light & X & R \\
  Refracted Light & & S \\
  Thermal Emissions & X & R \\
  Doppler Beaming & X & \\
  Ellipsoidal Variations & X & M \\
  Atmospheric Effects & & S \\
  Multiple Planets & & D \\
  Planetary Shapes & & S \\
  \bottomrule
 \end{tabular}
\label{tab:effects}
\end{center}
\end{table}

While it is true that by adding more parameters, one can over-fit the data by increasing the likelihood, for this reason, EXONEST employs Bayesian model testing, which enables one to evaluate the significance of each photometric effect. The model testing capabilities of EXONEST were demonstrated on Kepler Space Telescope of {Kepler}-76. {Kepler}-76b was demonstrated to exhibit several photometric effects: reflected light, Doppler boosting, ellipsoidal variations and~atmospheric superrotation. Parameter estimates were reported that are in agreement with previous studies~\cite{Faigler+etal:2013, Angerhausen+DeLarme+Morse:2015, Esteves+etal:2015:K91} using both radial velocity and photometric data. In addition, the model testing capabilities of EXONEST enable one to test photometric models against one another.

Future missions promise instruments of increasing sensitivity and detail with multiple spectral bands. Despite this, photometry will continue to be a valuable resource until detailed direct imaging is possible. In the meantime, we aim to include more relevant photometric effects into EXONEST and to develop the software package to work in conjunction with future multispectral studies.
\vspace{6pt}

\section*{Acknowledgements}
This paper was presented as an invited talk at the 37th International Workshop on Bayesian Inference and Maximum Entropy Methods in Science and Engineering (MaxEnt 2017) in Jarinu SP, Brasil. Knuth would like to thank the organizers, Adriano Polpo and Julio Stern, for both the kind invitation and their generous hospitality. We also wish to thank Jon Jenkins, Doug Caldwell, John Skilling, Paul Goggans, Wesley Henderson, Ariel Caticha, Keith Earle, Oleg Lunin and Matthew Szydagis for interesting discussions and helpful advice, questions and comments. We would also like to thank the two anonymous reviewers whose comments served to improve the quality of this manuscript.

\section*{Author Contributions}
K.H.K. and B.P. conceived of EXONEST. K.H.K. advised the research and work performed by B.P., J.L.C., B.D., A.D.G., and B.C., and wrote the greater part of the manuscript.  B.P. developed, coded, tested, and utilized EXONEST as a major component of his Ph.D. research.  B.P. performed the analysis of Kepler-76b and wrote the corresponding section of the manuscript.  D.A. contributed to the development of EXONEST and the inclusion of several photometric effects.  As the focus of her Ph.D. research, J.L.C. has led the effort to more accurately model reflected light, and she contributed to the writing of the section on reflected light.  As the focus of his M.S. thesis, B.D. implemented the three-body orbital code.  As the focus of his M.S. thesis, A.D.G. coded and tested the efficacy of stellar ellipsoidal variation models, and wrote the corresponding section on ellipsoidal variations.  B.C. assisted J.L.C. in the modeling of reflected light, and as the focus of his Ph.D. research, he has worked to characterize planetary shapes and their transit signatures.

\bibliographystyle{natbib}
\bibliography{knuth52}

\hyphenation{Post-Script Sprin-ger}
\begin{thebibliography}{}

\bibitem[Abraham and Shaw(1983)Abraham and Shaw]{Abraham+Shaw:1983:Dynamics}
Abraham, R.~H. and Shaw, C.~D. (1983).
\newblock {\em Dynamics-The Geometry of Behavior: Vol.: 2: Chaotic Behavior.}
\newblock Aerial Press, Incorporated.

\bibitem[{Angerhausen} {\em et~al.}(2015){Angerhausen}, {DeLarme}, and
  {Morse}]{Angerhausen+DeLarme+Morse:2015}
{Angerhausen}, D., {DeLarme}, E., and {Morse}, J.~A. (2015).
\newblock {A Comprehensive Study of {K}epler Phase Curves and Secondary
  Eclipses: Temperatures and Albedos of Confirmed {K}epler Giant Planets}.
\newblock {\em Publications of the Astronomical Society of the Pacific\/}, {\bf
  127}, 1113--1130.

\bibitem[Armstrong {\em et~al.}(2016)Armstrong, de~Mooij, Barstow, Osborn,
  Blake, and Saniee]{Armstrong+etal:2016}
Armstrong, D.~J., de~Mooij, E., Barstow, J., Osborn, H.~P., Blake, J., and
  Saniee, N.~F. (2016).
\newblock Variability in the atmosphere of the hot giant planet {HAT-P-7} b.
\newblock {\em Nature Astronomy\/}, {\bf 1}, 0004.

\bibitem[Artigau {\em et~al.}(2009)Artigau, Bouchard, Doyon, and
  Lafreni{\`e}re]{Artigau+etal:2009}
Artigau, {\'E}., Bouchard, S., Doyon, R., and Lafreni{\`e}re, D. (2009).
\newblock Photometric variability of the {T2}. 5 brown dwarf {SIMP J013656. 5+
  093347}: Evidence for evolving weather patterns.
\newblock {\em ApJ\/}, {\bf 701}(2), 1534.

\bibitem[Barnes(2007)Barnes]{Barnes:2007}
Barnes, J.~W. (2007).
\newblock Effects of orbital eccentricity on extrasolar planet transit
  detectability and light curves.
\newblock {\em Publications of the Astronomical Society of the Pacific\/}, {\bf
  119}(859), 986.

\bibitem[{Batygin} {\em et~al.}(2013){Batygin}, {Stanley}, and
  {Stevenson}]{Batygin+Stanley+Stevenson:2013}
{Batygin}, K., {Stanley}, S., and {Stevenson}, D.~J. (2013).
\newblock Magnetically controlled circulation on hot extrasolar planets.
\newblock {\em ApJ\/}, {\bf 776}, 53.

\bibitem[Beichman {\em et~al.}(2014)Beichman, Benneke, Knutson, Smith, Lagage,
  Dressing, Latham, Lunine, Birkmann, Ferruit, {\em
  et~al.}]{Beichman+etal:2014:JWST}
Beichman, C., Benneke, B., Knutson, H., Smith, R., Lagage, P.-O., Dressing, C.,
  Latham, D., Lunine, J., Birkmann, S., Ferruit, P., {\em et~al.} (2014).
\newblock Observations of transiting exoplanets with the {J}ames {W}ebb {S}pace
  {T}elescope ({JWST}).
\newblock {\em PASP\/}, {\bf 126}(946), 1134.

\bibitem[B\'{e}tr\'{e}mieux and Kaltenegger(2015)B\'{e}tr\'{e}mieux and
  Kaltenegger]{Betremieux+Kaltenegger:2015}
B\'{e}tr\'{e}mieux, Y. and Kaltenegger, L. (2015).
\newblock Refraction in planetary atmospheres: improved analytical expressions
  and comparison with a new ray-tracing algorithm.
\newblock {\em MNRAS\/}, {\bf 451}(2), 1268--1283.

\bibitem[{Borucki} and {Summers}(1984){Borucki} and
  {Summers}]{Borucki+Summers:1984}
{Borucki}, W.~J. and {Summers}, A.~L. (1984).
\newblock The photometric method of detecting other planetary systems.
\newblock {\em Icarus\/}, {\bf 58}, 121--134.

\bibitem[{Borucki} {\em et~al.}(2009){Borucki}, {Koch}, {Jenkins}, {Sasselov},
  {Gilliland}, {Batalha}, {Latham}, {Caldwell}, {Basri}, {Brown},
  {Christensen-Dalsgaard}, {Cochran}, {DeVore}, {Dunham}, {Dupree}, {Gautier},
  {Geary}, {Gould}, {Howell}, {Kjeldsen}, {Lissauer}, {Marcy}, {Meibom},
  {Morrison}, and {Tarter}]{Borucki-etal:2009:Science}
{Borucki}, W.~J., {Koch}, D., {Jenkins}, J., {Sasselov}, D., {Gilliland}, R.,
  {Batalha}, N., {Latham}, D.~W., {Caldwell}, D., {Basri}, G., {Brown}, T.,
  {Christensen-Dalsgaard}, J., {Cochran}, W.~D., {DeVore}, E., {Dunham}, E.,
  {Dupree}, A.~K., {Gautier}, T., {Geary}, J., {Gould}, A., {Howell}, S.,
  {Kjeldsen}, H., {Lissauer}, J., {Marcy}, G., {Meibom}, S., {Morrison}, D.,
  and {Tarter}, J. (2009).
\newblock {K}epler's optical phase curve of the exoplanet {HAT-P-7b}.
\newblock {\em Science\/}, {\bf 325}, 709.

\bibitem[{Borucki} {\em et~al.}(2010){Borucki}, {Koch}, {Basri}, {Batalha},
  {Brown}, {Caldwell}, {Caldwell}, {Christensen-Dalsgaard}, {Cochran},
  {DeVore}, {Dunham}, {Dupree}, {Gautier}, {Geary}, {Gilliland}, {Gould},
  {Howell}, {Jenkins}, {Kondo}, {Latham}, {Marcy}, {Meibom}, {Kjeldsen},
  {Lissauer}, {Monet}, {Morrison}, {Sasselov}, {Tarter}, {Boss}, {Brownlee},
  {Owen}, {Buzasi}, {Charbonneau}, {Doyle}, {Fortney}, {Ford}, {Holman},
  {Seager}, {Steffen}, {Welsh}, {Rowe}, {Anderson}, {Buchhave}, {Ciardi},
  {Walkowicz}, {Sherry}, {Horch}, {Isaacson}, {Everett}, {Fischer}, {Torres},
  {Johnson}, {Endl}, {MacQueen}, {Bryson}, {Dotson}, {Haas}, {Kolodziejczak},
  {Van Cleve}, {Chandrasekaran}, {Twicken}, {Quintana}, {Clarke}, {Allen},
  {Li}, {Wu}, {Tenenbaum}, {Verner}, {Bruhweiler}, {Barnes}, and
  {Prsa}]{Borucki:2010:Kepler}
{Borucki}, W.~J., {Koch}, D., {Basri}, G., {Batalha}, N., {Brown}, T.,
  {Caldwell}, D., {Caldwell}, J., {Christensen-Dalsgaard}, J., {Cochran},
  W.~D., {DeVore}, E., {Dunham}, E.~W., {Dupree}, A.~K., {Gautier}, T.~N.,
  {Geary}, J.~C., {Gilliland}, R., {Gould}, A., {Howell}, S.~B., {Jenkins},
  J.~M., {Kondo}, Y., {Latham}, D.~W., {Marcy}, G.~W., {Meibom}, S.,
  {Kjeldsen}, H., {Lissauer}, J.~J., {Monet}, D.~G., {Morrison}, D.,
  {Sasselov}, D., {Tarter}, J., {Boss}, A., {Brownlee}, D., {Owen}, T.,
  {Buzasi}, D., {Charbonneau}, D., {Doyle}, L., {Fortney}, J., {Ford}, E.~B.,
  {Holman}, M.~J., {Seager}, S., {Steffen}, J.~H., {Welsh}, W.~F., {Rowe}, J.,
  {Anderson}, H., {Buchhave}, L., {Ciardi}, D., {Walkowicz}, L., {Sherry}, W.,
  {Horch}, E., {Isaacson}, H., {Everett}, M.~E., {Fischer}, D., {Torres}, G.,
  {Johnson}, J.~A., {Endl}, M., {MacQueen}, P., {Bryson}, S.~T., {Dotson}, J.,
  {Haas}, M., {Kolodziejczak}, J., {Van Cleve}, J., {Chandrasekaran}, H.,
  {Twicken}, J.~D., {Quintana}, E.~V., {Clarke}, B.~D., {Allen}, C., {Li}, J.,
  {Wu}, H., {Tenenbaum}, P., {Verner}, E., {Bruhweiler}, F., {Barnes}, J., and
  {Prsa}, A. (2010).
\newblock {K}epler planet-detection mission: Introduction and first results.
\newblock {\em Science\/}, {\bf 327}, 977--.

\bibitem[Broeg {\em et~al.}(2013)Broeg, Fortier, Ehrenreich, Alibert,
  Baumjohann, Benz, Deleuil, Gillon, Ivanov, Liseau, Meyer, Oloffson, Pagano,
  Piotto, Pollacco, Queloz, Ragazzoni, Renotte, Steller, Thomas, and the
  {CHEOPS}~Team]{Broeg+etal:2013}
Broeg, C., Fortier, A., Ehrenreich, D., Alibert, Y., Baumjohann, W., Benz, W.,
  Deleuil, M., Gillon, M., Ivanov, A., Liseau, R., Meyer, M., Oloffson, G.,
  Pagano, I., Piotto, G., Pollacco, D., Queloz, D., Ragazzoni, R., Renotte, E.,
  Steller, M., Thomas, N., and the {CHEOPS}~Team (2013).
\newblock {CHEOPS} a transit photometry mission for {ESA}'s small mission
  programme.
\newblock {\em {EPJ} Web of Conferences\/}, {\bf 47}.

\bibitem[Brown(2009)Brown]{Brown:photometric:2009}
Brown, R.~A. (2009).
\newblock Photometric orbits of extrasolar planets.
\newblock {\em The Astrophysical Journal\/}, {\bf 702}(2), 1237.

\bibitem[Caltech(2017)Caltech]{ExoplanetArchive:2017}
Caltech (2017).
\newblock Nasa exoplanet archive.
\newblock \url{https://exoplanetarchive.ipac.caltech.edu/}.
\newblock Accessed: August 1, 2017.

\bibitem[Carter(ress)Carter]{Carter:thesis:2017}
Carter, J. (In Progress).
\newblock {\em Estimation of Planetary Photometric Emissions for Extremely
  Close-in Exoplanets\/}.
\newblock Ph.D. thesis, University at Albany.

\bibitem[Charbonneau {\em et~al.}(2005)Charbonneau, Allen, Megeath, Torres,
  Alonso, Brown, Gilliland, Latham, Mandushev, O'Donovan, {\em
  et~al.}]{Charbonneau+etal:2005}
Charbonneau, D., Allen, L.~E., Megeath, S.~T., Torres, G., Alonso, R., Brown,
  T.~M., Gilliland, R.~L., Latham, D.~W., Mandushev, G., O'Donovan, F.~T., {\em
  et~al.} (2005).
\newblock Detection of thermal emission from an extrasolar planet.
\newblock {\em ApJ\/}, {\bf 626}(1), 523.

\bibitem[{CHEOPS Science Team - University of Bern}(2017){CHEOPS Science Team -
  University of Bern}]{CHEOPSMHP}
{CHEOPS Science Team - University of Bern} (2017).
\newblock {CHEOPS} mission homepage.
\newblock http://cheops.unibe.ch/.
\newblock Accessed: 2017-03-28.

\bibitem[Dalba(2017)Dalba]{Dalba:2017}
Dalba, P.~A. (2017).
\newblock Out-of-transit refracted light in the atmospheres of transiting and
  non-transiting exoplanets.
\newblock {\em ApJ\/}, {\bf 848}(2), 91.

\bibitem[Demory {\em et~al.}(2013)Demory, De~Wit, Lewis, Fortney, Zsom, Seager,
  Knutson, Heng, Madhusudhan, Gillon, {\em et~al.}]{Demory+etal:2013:clouds}
Demory, B.-O., De~Wit, J., Lewis, N., Fortney, J., Zsom, A., Seager, S.,
  Knutson, H., Heng, K., Madhusudhan, N., Gillon, M., {\em et~al.} (2013).
\newblock Inference of inhomogeneous clouds in an exoplanet atmosphere.
\newblock {\em ApJ Lett\/}, {\bf 776}(2), L25.

\bibitem[ESA(2016a)ESA]{CHEOPS}
ESA (2016a).
\newblock {Science \& Technology:} {CHEOPS}.
\newblock http://sci.esa.int/cheops/.
\newblock Accessed: 2016-09-30.

\bibitem[ESA(2016b)ESA]{PLATO}
ESA (2016b).
\newblock Science \& technology: {PLATO}.
\newblock http://sci.esa.int/plato/.
\newblock Accessed: 2016-09-30.

\bibitem[{{ESA} Study Team} and {{PLATO} Science Study Team}(2010){{ESA} Study
  Team} and {{PLATO} Science Study Team}]{PlatoSciReq}
{{ESA} Study Team} and {{PLATO} Science Study Team} (2010).
\newblock {PLATO} science requirements document.
\newblock {\em ESA Public Documents\/}.

\bibitem[Esteves {\em et~al.}(2015)Esteves, De~Mooij, and
  Jayawardhana]{Esteves+etal:2015:K91}
Esteves, L., De~Mooij, E., and Jayawardhana, R. (2015).
\newblock Changing phases of alien worlds: Probing atmospheres of {K}epler
  planets with high-precision photometry.
\newblock {\em ApJ\/}, {\bf 804}(2), 150.

\bibitem[Esteves {\em et~al.}(2013)Esteves, De~Mooij, and
  Jayawardhana]{Esteves+DeMooij+Jayawardhana:2013}
Esteves, L.~J., De~Mooij, E.~J.~W., and Jayawardhana, R. (2013).
\newblock Optical phase curves of kepler exoplanets.
\newblock {\em ApJ\/}, {\bf 772}(1), 51.

\bibitem[{Faigler} and {Mazeh}(2007){Faigler} and {Mazeh}]{Faigler+Mazeh:2007}
{Faigler}, S. and {Mazeh}, T. (2007).
\newblock Photometric detection of non-transiting short-period low-mass
  companions through the beaming, ellipsoidal, and reflection effects in
  {Kepler} and {CoRoT} lightcurves.
\newblock {\em Monthly Notices of the Royal Astronomical Society\/}, {\bf
  1-19}.

\bibitem[{Faigler} and {Mazeh}(2015){Faigler} and {Mazeh}]{Faigler+Mazeh:2015}
{Faigler}, S. and {Mazeh}, T. (2015).
\newblock {BEER} analysis of {Kepler} and {CoRoT} lightcurves. {II}. evidence
  for superrotation in the phase curves of three {Kepler} hot jupiters.
\newblock {\em The Astrophysical Journal\/}, {\bf 800(1)}.

\bibitem[Faigler {\em et~al.}(2013)Faigler, Tal-Or, Mazeh, Latham, and
  Buchhave]{Faigler+etal:2013}
Faigler, S., Tal-Or, L., Mazeh, T., Latham, D.~W., and Buchhave, L.~A. (2013).
\newblock {BEER} analysis of {K}epler and {CoRoT} light curves. i. discovery of
  {K}epler-76b: A hot {J}upiter with evidence for superrotation.
\newblock {\em ApJ\/}, {\bf 771}(1), 26.

\bibitem[Feroz {\em et~al.}(2009)Feroz, Hobson, and Bridges]{Feroz+etal:2009}
Feroz, F., Hobson, M.~P., and Bridges, M. (2009).
\newblock {M}ulti{N}est: an efficient and robust {B}ayesian inference tool for
  cosmology and particle physics.
\newblock {\em MNRAS\/}, {\bf 398}(4), 1601--1614.

\bibitem[Feroz {\em et~al.}(2011)Feroz, Balan, and Hobson]{Feroz+etal:2011a}
Feroz, F., Balan, S.~T., and Hobson, M.~P. (2011).
\newblock Detecting extrasolar planets from stellar radial velocities using
  {B}ayesian evidence.
\newblock {\em MNRAS\/}, {\bf 415}(4), 3462--3472.

\bibitem[Feroz {\em et~al.}(2013)Feroz, Hobson, Cameron, and
  Pettitt]{Feroz+etal:2013}
Feroz, F., Hobson, M.~P., Cameron, E., and Pettitt, A.~N. (2013).
\newblock Importance nested sampling and the multinest algorithm.
\newblock {\em arXiv preprint arXiv:1306.2144\/}.

\bibitem[Ford(2006)Ford]{Ford:2006:MCMC}
Ford, E.~B. (2006).
\newblock Improving the efficiency of {M}arkov {C}hain {M}onte {C}arlo for
  analyzing the orbits of extrasolar planets.
\newblock {\em ApJ\/}, {\bf 642}(1), 505.

\bibitem[Ford {\em et~al.}(2001)Ford, Seager, and
  Turner]{Ford+Seager+Turner:2001}
Ford, E.~B., Seager, S., and Turner, E.~L. (2001).
\newblock Characterization of extrasolar terrestrial planets from diurnal
  photometric variability.
\newblock {\em arXiv preprint astro-ph/0109054\/}.

\bibitem[Gai(2016)Gai]{Gai:thesis:2016}
Gai, A. (2016).
\newblock {\em Bayesian Model Testing of Models for Ellipsoidal Variation on
  Stars Due to Hot {J}upiters\/}.
\newblock Master's thesis, University at Albany.

\bibitem[Gai and Knuth(2017)Gai and Knuth]{Gai+Knuth:2017}
Gai, A. and Knuth, K.~H. (2017).
\newblock Bayesian model testing of ellipsoidal variations on stars due to hot
  {J}upiters.
\newblock In Prep.

\bibitem[Gillon {\em et~al.}(2013)Gillon, Triaud, Jehin, Delrez, Opitom,
  Magain, Lendl, and Queloz]{Gillon+etal:2013}
Gillon, M., Triaud, A.~H.~M.~J., Jehin, E., Delrez, L., Opitom, C., Magain, P.,
  Lendl, M., and Queloz, D. (2013).
\newblock Fast-evolving weather for the coolest of our two new substellar
  neighbours.
\newblock {\em Astronomy \& Astrophysics\/}, {\bf 555}, L5.

\bibitem[Gregersen(2010)Gregersen]{Gregersen:2010}
Gregersen, E. (2010).
\newblock {\em The {M}ilky {W}ay and Beyond: Stars, Nebulae, and Other
  Galaxies\/}.
\newblock The Rosen Publishing Group.

\bibitem[Hastings(1970)Hastings]{Hastings:1970}
Hastings, W.~K. (1970).
\newblock {M}onte {C}arlo sampling methods using {M}arkov chains and their
  applications.
\newblock {\em Biometrika\/}, {\bf 57}(1), 97–109.

\bibitem[Heller and Pudritz(2016)Heller and Pudritz]{Heller+Pudritz:2016}
Heller, R. and Pudritz, R.~E. (2016).
\newblock The search for extraterrestrial intelligence in {E}arth's solar
  transit zone.
\newblock {\em Astrobiology\/}, {\bf 16}(4), 259--270.

\bibitem[Heng(2012)Heng]{Heng:2012:shocks}
Heng, K. (2012).
\newblock On the existence of shocks in irradiated exoplanetary atmospheres.
\newblock {\em ApJ Lett\/}, {\bf 761}(1), L1.

\bibitem[Heng and Demory(2013)Heng and Demory]{Heng+Demory:2013}
Heng, K. and Demory, B.-O. (2013).
\newblock Understanding trends associated with clouds in irradiated exoplanets.
\newblock {\em ApJ\/}, {\bf 777}(2), 100.

\bibitem[Heyrovsk\'{y}(2007)Heyrovsk\'{y}]{Heyrovsky:2007}
Heyrovsk\'{y}, D. (2007).
\newblock Computing limb-darkening coefficients from stellar atmosphere models.
\newblock {\em The Astrophysical Journal\/}, {\bf 656(1)}, 483--492.

\bibitem[Hippke and Angerhausen(2015)Hippke and
  Angerhausen]{Hippke+Angerhausen:2015}
Hippke, M. and Angerhausen, D. (2015).
\newblock Photometry's bright future: Detecting solar system analogs with
  future space telescopes.
\newblock {\em ApJ\/}, {\bf 810}(1), 29.

\bibitem[Jackson {\em et~al.}(2012)Jackson, Lewis, Barnes, Deming, Showman, and
  Fortney]{Jackson+etal:2012}
Jackson, B.~K., Lewis, N., Barnes, J., Deming, L., Showman, A., and Fortney, J.
  (2012).
\newblock The {EVIL-MC} model for ellipsoidal variations of planet-hosting
  stars and applications to the {HAT-P-7} system.
\newblock {\em The Astrophysical Journal\/}, {\bf 751:112 (13pp)}.

\bibitem[Jaynes(2003)Jaynes]{Jaynes:Book}
Jaynes, E.~T. (2003).
\newblock {\em Probability Theory: The Logic of Science\/}.
\newblock Cambridge Univ. Press, Cambridge.

\bibitem[Jenkins and Doyle(2003)Jenkins and Doyle]{Jenkins+Doyle:2003}
Jenkins, J.~M. and Doyle, L.~R. (2003).
\newblock Detecting reflected light from close-in extrasolar giant planets with
  the {K}epler photometer.
\newblock {\em The Astrophysical Journal\/}, {\bf 595}(1), 429--445.

\bibitem[Kane and Gelino(2012)Kane and Gelino]{Kane+Gelino:2012}
Kane, S.~R. and Gelino, D.~M. (2012).
\newblock Distinguishing between stellar and planetary companions with phase
  monitoring.
\newblock {\em MNRAS\/}, {\bf 424(1)}, 779–788.

\bibitem[Knuth {\em et~al.}(2015)Knuth, Habeck, Malakar, Mubeen, and
  Placek]{Knuth+etal:DSP:2015}
Knuth, K.~H., Habeck, M., Malakar, N.~K., Mubeen, A.~M., and Placek, B. (2015).
\newblock {B}ayesian evidence and model selection.
\newblock {\em Digital Signal Processing\/}, {\bf 47}, 50--67.
\newblock (arXiv:1411.3013).

\bibitem[{\'K}opal(1954){\'K}opal]{Kopal:1953}
{\'K}opal, Z. (1954).
\newblock Photometric effects of reflection in close binary systems.
\newblock {\em MNRAS\/}, {\bf 114}(1), 101--117.

\bibitem[{\'K}opal(1959){\'K}opal]{Kopal:1959}
{\'K}opal, Z. (1959).
\newblock {\em Close Binary Systems\/}.
\newblock New York John Wiley \& Sons Inc.

\bibitem[L{\'e}ger {\em et~al.}(2011)L{\'e}ger, Grasset, Fegley, Codron,
  Albarede, Barge, Barnes, Cance, Carpy, Catalano, {\em
  et~al.}]{Leger+etal:2011}
L{\'e}ger, A., Grasset, O., Fegley, B., Codron, F., Albarede, A.~F., Barge, P.,
  Barnes, R., Cance, P., Carpy, S., Catalano, F., {\em et~al.} (2011).
\newblock The extreme physical properties of the {CoRoT}-7b super-{E}arth.
\newblock {\em Icarus\/}, {\bf 213}(1), 1--11.

\bibitem[Lock and Stewart(2017)Lock and Stewart]{Lock+Stewart:2017:synestias}
Lock, S.~J. and Stewart, S.~T. (2017).
\newblock The structure of terrestrial bodies: Impact heating, corotation
  limits, and synestias.
\newblock {\em Journal of Geophysical Research: Planets\/}, {\bf 122}(5),
  950–--982.
\newblock (arXiv:1705.07858 [astro-ph.EP]).

\bibitem[Mandel and Agol(2002)Mandel and Agol]{Mandel+Agol:2002}
Mandel, K. and Agol, E. (2002).
\newblock Analytic light curves for planetary transit searches.
\newblock {\em ApJ Lett\/}, {\bf 580}(2), L171.

\bibitem[Maurin {\em et~al.}(2012)Maurin, Selsis, Hersant, and
  Belu]{Maurin+etal:2012}
Maurin, A.~S., Selsis, F., Hersant, F., and Belu, A. (2012).
\newblock Thermal phase curves of nontransiting terrestrial exoplanets-ii.
  characterizing airless planets.
\newblock {\em Astronomy \& Astrophysics\/}, {\bf 538}, A95.

\bibitem[Menou(2012)Menou]{Menou:2012}
Menou, K. (2012).
\newblock Magnetic scaling laws for the atmospheres of hot giant exoplanets.
\newblock {\em ApJ\/}, {\bf 745}(2), 138.

\bibitem[{Metropolis} {\em et~al.}(1953){Metropolis}, {Rosenbluth},
  {Rosenbluth}, {Teller}, and {Teller}]{Metropolis:1953}
{Metropolis}, N., {Rosenbluth}, A.~W., {Rosenbluth}, M.~N., {Teller}, A.~H.,
  and {Teller}, E. (1953).
\newblock Equations of state calculations by fast computing machines.
\newblock {\em Journal of Chemical Physics\/}, {\bf 21}(6), 1087--1092.

\bibitem[Misra(2014)Misra]{Misra:2014}
Misra, A. (2014).
\newblock {\em The Effects of Refraction and Forward Scattering on Exoplanet
  Transit Transmission Spectroscopy\/}.
\newblock Ph.D. thesis, University of Washington.

\bibitem[Misra {\em et~al.}(2014)Misra, Meadows, and
  Crisp]{Misra+Meadows+Crisp:2014}
Misra, A., Meadows, V., and Crisp, D. (2014).
\newblock The effects of refraction on transit transmission spectroscopy:
  application to {E}arth-like exoplanets.
\newblock {\em The Astrophysical Journal\/}, {\bf 792}(1), 61.

\bibitem[Morris(1985)Morris]{Morris:1985}
Morris, S.~L. (1985).
\newblock The ellipsoidal variable stars.
\newblock {\em The Astrophysical Journal\/}, {\bf 295}, 143--152.

\bibitem[Mu{\~n}oz and Cabrera(2017)Mu{\~n}oz and Cabrera]{Munoz+Cabrera:2017}
Mu{\~n}oz, A.~G. and Cabrera, J. (2017).
\newblock Exoplanet phase curves at large phase angles. {D}iagnostics for
  extended hazy atmospheres.
\newblock {\em MNRAS\/}.
\newblock In Press.

\bibitem[Mu{\~n}oz {\em et~al.}(2012)Mu{\~n}oz, Osorio, Barrena,
  Monta{\~n}{\'e}s-Rodr{\'\i}guez, Mart{\'\i}n, and Pall{\'e}]{Munoz+etal:2012}
Mu{\~n}oz, A.~G., Osorio, M.~R.~Z., Barrena, R.,
  Monta{\~n}{\'e}s-Rodr{\'\i}guez, P., Mart{\'\i}n, E.~L., and Pall{\'e}, E.
  (2012).
\newblock Glancing views of the {E}arth: from a lunar eclipse to an
  exoplanetary transit.
\newblock {\em ApJ\/}, {\bf 755}(2), 103.

\bibitem[Mu{\~n}oz {\em et~al.}(2017)Mu{\~n}oz, Lavvas, and
  West]{Munoz+Lavvas+West:2017:Titan}
Mu{\~n}oz, A.~G., Lavvas, P., and West, R.~A. (2017).
\newblock {T}itan brighter at twilight than in daylight.
\newblock {\em Nature Astronomy\/}, {\bf 1}, 0114.
\newblock arXiv:1704.07460 [astro-ph.EP].

\bibitem[NASA(2014)NASA]{NASA:2014}
NASA (2014).
\newblock {\em NASA Strategic Plan 2014\/}.
\newblock NASA.

\bibitem[NASA(2016)NASA]{WFIRST}
NASA (2016).
\newblock Wide field infrared survey telescope.
\newblock https://wfirst.gsfc.nasa.gov/index.html.
\newblock Accessed: 2016-10-29.

\bibitem[Pall{\'e} {\em et~al.}(2008)Pall{\'e}, Ford, Seager,
  Monta{\~n}{\'e}s-Rodr{\'\i}guez, and Vazquez]{Palle+etal:2008}
Pall{\'e}, E., Ford, E.~B., Seager, ., Monta{\~n}{\'e}s-Rodr{\'\i}guez, P., and
  Vazquez, M. (2008).
\newblock Identifying the rotation rate and the presence of dynamic weather on
  extrasolar {E}arth-like planets from photometric observations.
\newblock {\em ApJ\/}, {\bf 676}(2), 1319.

\bibitem[Perryman(2011)Perryman]{Perryman:2011}
Perryman, M. (2011).
\newblock {\em The exoplanet handbook\/}.
\newblock Cambridge University Press.

\bibitem[Placek(2014)Placek]{Placek:thesis:2014}
Placek, B. (2014).
\newblock {\em {B}ayesian detection and characterization of extra-solar planets
  via photometric variations\/}.
\newblock Ph.D. thesis, University at Albany (SUNY).

\bibitem[Placek {\em et~al.}(2014)Placek, Knuth, and
  Angerhausen]{Placek+Knuth+Angerhausen:EXONEST}
Placek, B., Knuth, K.~H., and Angerhausen, D. (2014).
\newblock {EXONEST}: {B}ayesian model selection applied to the detection and
  characterization of exoplanets via photometric variations.
\newblock {\em Astrophys. J.}, {\bf 795}(2), 112.
\newblock arXiv:1310.6764 [astro-ph.EP].

\bibitem[Placek {\em et~al.}(2015)Placek, Knuth, Angerhausen, and
  Jenkins]{Placek+etal:K91:2015}
Placek, B., Knuth, K., Angerhausen, D., and Jenkins, J. (2015).
\newblock Characterization of {K}epler-91b and the investigation of a potential
  trojan companion using {EXONEST}.
\newblock {\em The Astrophysical Journal\/}, {\bf 814}(2), 147.

\bibitem[Placek {\em et~al.}(2016)Placek, Knuth, and
  Angerhausen]{Placek+Knuth+Angerhausen:2016:Kepler+TESS}
Placek, B., Knuth, K.~H., and Angerhausen, D. (2016).
\newblock Combining photometry from {K}epler and {TESS} to improve short-period
  exoplanet characterization.
\newblock {\em PASP\/}, {\bf 128}(7), 074503.

\bibitem[Placek {\em et~al.}(2017)Placek, Angerhausen, and
  Knuth]{Placek+Angerhausen+Knuth:2017:optimization}
Placek, B., Angerhausen, D., and Knuth, K.~H. (2017).
\newblock Optimizing observing strategies for exoplanet secondary eclipses and
  phase curves.
\newblock {\em ApJ\/}.
\newblock In Press.

\bibitem[{Rauscher} and {Menou}(2013){Rauscher} and
  {Menou}]{Rauscher+Menou:2013}
{Rauscher}, E. and {Menou}, K. (2013).
\newblock Three-dimensional atmospheric circulation models of {HD} 189733b and
  {HD} 209458b with consistent magnetic drag and ohmic dissipation.
\newblock {\em ApJ\/}, {\bf 764}, 103.

\bibitem[Ricker {\em et~al.}(2010)Ricker, Latham, Vanderspek, Ennico, Bakos,
  Brown, Burgasser, Charbonneau, Clampin, Deming, {\em
  et~al.}]{Ricker+etal:2010:TESS}
Ricker, G.~R., Latham, D.~W., Vanderspek, R.~K., Ennico, K.~A., Bakos, G.,
  Brown, T.~M., Burgasser, A.~J., Charbonneau, D., Clampin, M., Deming, L.~D.,
  {\em et~al.} (2010).
\newblock Transiting exoplanet survey satellite (tess).
\newblock In {\em Bulletin of the American Astronomical Society\/}, volume~42,
  page 459.

\bibitem[Rybicki and Lightman(2008)Rybicki and Lightman]{Rybicki+Lightman:2008}
Rybicki, G.~B. and Lightman, A.~P. (2008).
\newblock {\em Radiative processes in astrophysics\/}.
\newblock John Wiley \& Sons.

\bibitem[Seager(2010)Seager]{Seager:2010}
Seager, S. (2010).
\newblock {\em Exoplanet atmospheres: physical processes\/}.
\newblock Princeton University Press.

\bibitem[Seager and Mall{\v e}n-Ornelas(2003)Seager and Mall{\v
  e}n-Ornelas]{Seager+Mallen:2003}
Seager, S. and Mall{\v e}n-Ornelas, G. (2003).
\newblock A unique solution of planet and star parameters from an extrasolar
  planet transit light curve.
\newblock {\em ApJ\/}, {\bf 585}(2), 1038.

\bibitem[Showman and Polvani(2011)Showman and
  Polvani]{Showman+Polvani:2011:superrotation}
Showman, A.~P. and Polvani, L.~M. (2011).
\newblock Equatorial superrotation on tidally locked exoplanets.
\newblock {\em ApJ\/}, {\bf 738}(1), 71.

\bibitem[Sidis(2010)Sidis]{Sidis:2010}
Sidis, O. (2010).
\newblock Transits of transparent planets—atmospheric lensing effects.
\newblock {\em The Astrophysical Journal\/}, {\bf 720}(1), 904.

\bibitem[Sing(2010)Sing]{Sing:2010}
Sing, D. (2010).
\newblock Stellar limb-darkening coefficients for {CoRoT} and {Kepler}.
\newblock {\em Astronomy \& Astrophysics\/}, {\bf 510}(1), A21.

\bibitem[Sivia and Skilling(2006)Sivia and Skilling]{Sivia&Skilling}
Sivia, D.~S. and Skilling, J. (2006).
\newblock {\em Data Analysis. A {B}ayesian Tutorial\/}.
\newblock Oxford University Press, Oxford, second edition.

\bibitem[Skilling(2004)Skilling]{Skilling:2004:nested}
Skilling, J. (2004).
\newblock Nested sampling.
\newblock In R.~Fischer, V.~Dose, R.~Preuss, and U.~von Toussaint, editors,
  {\em Bayesian Inference and Maximum Entropy Methods in Science and
  Engineering, Garching, Germany 2004\/}, number 735 in AIP Conf. Proc., pages
  395--405. AIP, New York.

\bibitem[Skilling(2006)Skilling]{Skilling:2006:nested}
Skilling, J. (2006).
\newblock Nested sampling for general {B}ayesian computation.
\newblock {\em Bayesian Analysis\/}, {\bf 1}(4), 833--859.

\bibitem[Sliski and Kipping(2014)Sliski and Kipping]{Sliski+Kipping:2014}
Sliski, D.~H. and Kipping, D.~M. (2014).
\newblock A high false positive rate for {K}epler planetary candidates of giant
  stars using asterodensity profiling.
\newblock {\em ApJ\/}, {\bf 788}(2), 148.

\bibitem[Sobolev(1975)Sobolev]{Sobolev:1975}
Sobolev, V.~V. (1975).
\newblock {\em Light Scattering in Planetary Atmospheres\/}.
\newblock Oxford: Pergamon.

\bibitem[Van~Cleve and Caldwell(2009)Van~Cleve and Caldwell]{KeplerHandbook}
Van~Cleve, J.~E. and Caldwell, D.~A. (2009).
\newblock {\em Kepler Instrument Handbook\/}.
\newblock NASA/Ames Research Center, Moffett Field, California.

\bibitem[Van~Eylen {\em et~al.}(2013)Van~Eylen, Nielsen, Hinrup, Tingley, and
  Kjeldsen]{VanEylen+etal:2013}
Van~Eylen, V., Nielsen, M.~L., Hinrup, B., Tingley, B., and Kjeldsen, H.
  (2013).
\newblock Investigation of systematic effects in kepler data: Seasonal
  variations in the light curve of {HAT-P-7b}.
\newblock {\em ApJ Lett\/}, {\bf 774}(2), L19.

\end{thebibliography}

\end{document}